\documentclass[12pt,letterpaper,a4paper]{article}
\usepackage[includeheadfoot,
marginratio={1:1,2:3},
width=500pt,
height=730pt,]{geometry}
\usepackage{amsfonts}
\usepackage{amssymb}
\usepackage{comment}
\usepackage{graphicx}
\usepackage{amsmath}
\usepackage{tabu}
\usepackage{dsfont}
\usepackage{float}
\usepackage{amsthm}
\usepackage{slashed}
\usepackage{setspace}
\usepackage[labelformat=simple]{subcaption}

\usepackage{amsmath, amsthm, amssymb, amsfonts}
\usepackage[font=small, labelfont=bf]{caption}
\usepackage{amsmath, amsthm, amssymb, amsfonts}
\usepackage{multirow}
\usepackage{graphicx}
\usepackage{float}
\usepackage[numbers,sort&compress]{natbib}
\usepackage{enumerate}

\usepackage{amsmath}
\usepackage{amsfonts}
\usepackage{amssymb}
\usepackage{graphicx}
\usepackage{cancel}

\usepackage{empheq}
\usepackage{lipsum}
\usepackage{xpatch}
\usepackage{color}
\usepackage{hyperref}

\usepackage{float}
\usepackage{framed}
\usepackage[framemethod=tikz]{mdframed}
\restylefloat{table}
\restylefloat{figure}

\newcommand{\nc}{\newcommand}
\nc{\beq}{\begin{equation}}
\nc{\eeq}{\end{equation}}
\nc{\bea}{\begin{eqnarray}}
\nc{\eea}{\end{eqnarray}}

\def\ov{\overline}

\allowdisplaybreaks[2]
\numberwithin{equation}{section}

\def\rmt{\rm t}
\def\rmz{\rm z}

\begin{document}
	{\hfill
		%
		arXiv:1909.08630}
	
	\vspace{1.0cm}
	\begin{center}
		{\Large
			$T$-dualizing the de-Sitter no-go scenarios}
		\vspace{0.4cm}
	\end{center}
	
	\vspace{0.35cm}
	\begin{center}
		Pramod Shukla\footnote{Email: pramodmaths@gmail.com}
	\end{center}

	\vspace{0.1cm}
	\begin{center}
		{ICTP, Strada Costiera 11, Trieste 34151, Italy.}
	\end{center}
	
	\vspace{1cm}
	
	\abstract{In the context of realizing de-Sitter vacua and the slow-roll inflation, several no-go conditions have been found in the framework of type IIA (generalized) flux compactifications. In this article, using our recently proposed $T$-dual dictionary in \cite{Shukla:2019wfo}, we translate various such type IIA no-go conditions which subsequently leads to some interesting de-Sitter no-go scenarios in the presence of (non-)geometric fluxes on the dual type IIB side. We also present the relevance of using $K3/{\mathbb T}^4$-fibred Calabi Yau threefolds in order to facilitate one particular class of the de-Sitter no-go conditions. This analysis helps in refining certain corners of the vast non-geometric flux landscape for the hunt of de-Sitter vacua.}

\clearpage
	
\tableofcontents
	
\section{Introduction}
\label{sec_intro}
Recent revival of the swampland conjecture \cite{Obied:2018sgi, Agrawal:2018own} has boosted a huge amount of interest towards exploring the (non-)existence of de-Sitter vacua within a consistent theory of quantum gravity. The original idea of swampland has been proposed to state that de-Sitter solutions must be absent in a consistent theory of quantum gravity \cite{Ooguri:2006in}. This idea has been recently endorsed as a bound involving the scalar potential ($V$) and its derivatives given in the following manner,
\bea
\label{eq:old-swamp}
& & \frac{|\nabla V|}{V} \geq \frac{c}{M_p} \,, 
\eea
where the constant $c$ is an order one quantity. This conjecture has been supported by several explicit computations in the context of attempts made for realizing classical de-Sitter solutions and inflationary cosmology in the type II superstring flux compactifications \cite{Maldacena:2000mw, Hertzberg:2007wc, Hertzberg:2007ke, Haque:2008jz, Flauger:2008ad, Caviezel:2008tf,  Covi:2008ea, deCarlos:2009fq, Caviezel:2009tu, Danielsson:2009ff, Danielsson:2010bc, Wrase:2010ew,  Shiu:2011zt, McOrist:2012yc, Dasgupta:2014pma, Gautason:2015tig, Junghans:2016uvg, Andriot:2016xvq, Andriot:2017jhf, Danielsson:2018ztv}.  Note that, the bound presented in eqn. (\ref{eq:old-swamp}) does not only forbid the de-Sitter minima but also the de-Sitter maxima as well, and several counter examples were known \cite{Haque:2008jz, Danielsson:2009ff, Danielsson:2011au, Chen:2011ac, Danielsson:2012et} or have been reported soon after the proposal was made \cite{Andriot:2018wzk, Andriot:2018ept, Garg:2018reu, Denef:2018etk, Conlon:2018eyr, Roupec:2018mbn, Murayama:2018lie, Choi:2018rze, Hamaguchi:2018vtv, Olguin-Tejo:2018pfq, Blanco-Pillado:2018xyn} reflecting the need of refining the de-Sitter swampland conjecture in eqn. (\ref{eq:old-swamp}). Subsequently a refined version of the conjecture has been proposed which states that at least one of the following two constraints should always hold \cite{Ooguri:2018wrx},
\bea
\label{eq:new-swamp}
& & \frac{|\nabla V|}{V} \geq \frac{c}{M_p}\, , \qquad \qquad {\rm min} \Bigl[\frac{\nabla_i \nabla_j V}{V}\Bigr] \leq  - \frac{c'}{M_p^2}\,,
\eea
where $c$ and $c' > 0$ are order one constants. Note that these two parameters can be related to the usual inflationary parameters, namely the $\epsilon_V$ and $\eta_V$ parameters, which are needed to be sufficiently small for having the slow-roll inflation (e.g. see \cite{BlancoPillado:2006he, Hertzberg:2007ke,Hetz:2016ics}),
\bea
& & \hskip-1cm \epsilon_V \geq \frac{1}{2}\, c^2 \, , \qquad \qquad |\eta_V| \leq c' \,.
\eea
Therefore it is rather quite obvious that the conjecture (\ref{eq:new-swamp}) poses an obstruction to not only realising de-Sitter vacua but also in realising slow-roll inflationary scenarios, which demands $\epsilon_V \ll 1$ and $|\eta_V| \ll 1$. However, this definition of the $\epsilon_V$ parameter follows from a more general definition given in terms of Hubble parameter as $\epsilon_H = - \dot{H}/H^2$ which only needs to satisfy $\epsilon_H < 1$ for having an accelerated universe. This leads to a possible window circumventing the conjecture in the multi-field inflation with turning trajectories \cite{Hetz:2016ics,Achucarro:2018vey}. Moreover given the fact that no universal theoretical quantification of the $c$ and $c'$ parameters being available (though some experimental estimates have been reported in \cite{Raveri:2018ddi}), the order one statement may still keep some window open \cite{Kehagias:2018uem, Kinney:2018kew}. 

The question of realising de-Sitters is two-fold; first is about its existence and the second is about the stability, and a plethora of interesting models have been proposed on these lines \cite{Maldacena:2000mw, Hertzberg:2007wc, Caviezel:2008tf,Flauger:2008ad, Danielsson:2009ff, Caviezel:2009tu, Wrase:2010ew, Covi:2008ea, Shiu:2011zt, Danielsson:2012et, Junghans:2016uvg, Banlaki:2018ayh}. The swampland conjecture \cite{Ooguri:2006in} has been also found to be in close connections with the allowed inflaton field range in a trustworthy effective field description as it has been argued that massive tower of states can get excited after a certain limit to the inflaton excursions \cite{Blumenhagen:2017cxt, Blumenhagen:2018nts, Blumenhagen:2018hsh, Palti:2017elp, Conlon:2016aea, Hebecker:2017lxm, Klaewer:2016kiy, Baume:2016psm, Landete:2018kqf, Cicoli:2018tcq, Font:2019cxq, Grimm:2018cpv, Hebecker:2018fln, Banlaki:2018ayh, Junghans:2018gdb}. The recent surge of developments following the recent swampland proposal can be found in \cite{Denef:2018etk, Conlon:2018eyr, Garg:2018reu, Kinney:2018nny, Roupec:2018mbn, Murayama:2018lie, Choi:2018rze, Hamaguchi:2018vtv, Olguin-Tejo:2018pfq, Blanco-Pillado:2018xyn, Achucarro:2018vey, Kehagias:2018uem, Kinney:2018kew, Andriot:2018wzk, Andriot:2018ept, Lin:2018kjm, Han:2018yrk, Raveri:2018ddi, Dasgupta:2018rtp, Danielsson:2018qpa, Andriolo:2018yrz, Dasgupta:2019gcd, Russo:2018akp, Russo:2019fnk, Andriot:2019wrs} with an extensive review on the status in \cite{Palti:2019pca}.

In contrary to the (minimal) de-Sitter no-go scenarios, there have been several proposals for realizing stable de-Sitter vacua in the context of string model building \cite{Kachru:2003aw, Burgess:2003ic, Achucarro:2006zf, Westphal:2006tn, Silverstein:2007ac, Rummel:2011cd, Cicoli:2012fh, Louis:2012nb, Cicoli:2013cha, Cicoli:2015ylx, Cicoli:2017shd, Akrami:2018ylq, Antoniadis:2018hqy}; see \cite{Heckman:2019dsj, Heckman:2018mxl} also for the $F$-theoretic initiatives taken in this regard. In fact realizing de-Sitter solutions and possible obstructions on the way of doing it have been always in the center of attraction since decades\footnote{For an updated recent review on realizing de-Sitter solutions in string theoretic models along with the status on Quintessence, we refer the readers to \cite{Cicoli:2018kdo}.}. Moreover, some interesting models realizing de-Sitter vacua in the framework of non-geometric flux compactifications have also been proposed \cite{deCarlos:2009qm, deCarlos:2009fq, Danielsson:2010bc, Danielsson:2012by, Blaback:2013ht, Damian:2013dq, Damian:2013dwa, Blaback:2013fca, Hassler:2014mla, Blumenhagen:2015xpa, Blaback:2018hdo, Damian:2018tlf}. However the issues related to fluxes being integral and whether they satisfy all the NS-NS Bianchi identities or not, can still be considered to be among some open questions in this regard. In fact it has been observed that the Bianchi identities are not fully known in the beyond toroidal examples as there have been some inconsistencies observed in two ways of deriving the identities \cite{Ihl:2007ah, Robbins:2007yv, Gao:2018ayp, Shukla:2016xdy, Shukla:2019akv}. 

\subsubsection*{Motivations, goals and a brief summary of the results:}
Several de-Sitter No-Go theorems on the type IIA side have been well known since a decade or so \cite{Hertzberg:2007wc, Flauger:2008ad, Wrase:2010ew, Hertzberg:2007ke}, which have been also studied for the type II non-geometric compactifications using a simple isotropic torus case in \cite{deCarlos:2009qm, deCarlos:2009fq}. With a goal of extending the non-geometric flux phenomenology beyond the toroidal cases, the study of generic four-dimensional type II scalar potentials and their ten-dimensional origin have been performed in a series of papers \cite{Shukla:2015rua, Shukla:2015hpa, Blumenhagen:2015lta, Blumenhagen:2015kja, Shukla:2016hyy, Gao:2017gxk}. Taking this programme to one step further, in a companion paper \cite{Shukla:2019wfo}, we have presented a one-to-one $T$-dual mapping of the two type II effective scalar potentials, along with the flux constraints arising from the NS-NS Bianchi identities and the tadpole cancellation conditions, which are also in one-to-one correspondence under $T$-duality. The main motivations and the goals aimed in this article can be presented under the following points:
\begin{itemize}
\item{Our so-called ``cohomology" or ``symplectic" formulation of the scalar potential presented in \cite{Shukla:2019wfo} opens up the window to study the non-geometric models beyond the toroidal constructions, and also enables one to explicitly translate any useful findings of one setup into its $T$-dual picture. On these lines, we plan to $T$-dualize the several de-Sitter no-go scenarios realized in some purely geometric type IIA frameworks \cite{Hertzberg:2007wc, Flauger:2008ad, Wrase:2010ew, Hertzberg:2007ke}. This helps us in delving into their type IIB counterparts which turn out to be non-geometric de-Sitter no-go frameworks, and those have not been known before. 
The utility of our approach can be underlined by the fact that although the type IIA no-go scenarios have been known since more than a decade, there have been no de-Sitter no-go proposals in generic non-geometric type IIB framework. 
}
\item{In our analysis, we show the relevance of considering the complex structure moduli in deriving the $T$-dual type IIB no-go conditions. Note that all the type IIA no-go results in  \cite{Hertzberg:2007wc, Flauger:2008ad, Wrase:2010ew, Hertzberg:2007ke}, which we $T$-dualize, are realized using the extremization conditions only in the `volume/dilaton" plane, and without taking into account the complex structure moduli sector. This illustrates that any claim of evading the no-go originated from the ``volume/dilaton" analysis should be checked by including all the remaining moduli.}
\item{On the lines of classifying type IIA and type IIB models based on their (non-)geometric nature via turning on certain set of fluxes at a time, we present an interesting recipe which corresponds to considering what we call some `special solutions' of the NS-NS Bianchi identities. These solutions are such that they lead to a purely geometric framework as a $T$-dual of a non-geometric setup on either of the respective IIA or IIB sides. In particular, type IIA non-geometric model with fluxes allowed as in the `special solution' of the Bianchi identities is $T$-dual to a purely geometric type IIB model, which has been known to have de-Sitter no-go scenario \cite{Shiu:2011zt, Garg:2018reu}, and subsequently our analysis concludes that the corresponding $T$-dual type IIA model despite having the non-geometric fluxes (still allowed by the `special solution') cannot evade the no-go result. 
	
This shows that our approach will be useful for playing with constructing models in search of the de-Sitter no-go or against those no-go arguments, given that the most generic non-geometric setup could still be expected to evade the no-go, though there are several specifics to be checked in a given model before arriving at any final conclusion.}
\item{In addition to finding the (non-)geometric flux-regime or the types of fluxes needed to evade a certain kind of de-Sitter no-go result, we also find that if there are some specific geometries involved, such as $K3/{\mathbb T}^4$-fibred complex threefold, then there can be a restoration of the no-go results despite the inclusion of those fluxes which apparently could be anticipated to evade the respective no-go results. We illustrate this observation for explicit type IIA and IIB toroidal non-geometric setups.}
\end{itemize}
So, our results can be considered as providing some systematics about constructing de-Sitter no-go scenarios along with the recipes to find the possibilities of evading them, and at the same time, in looking for some specific geometries of the moduli space which could again restore the de-Sitter no-go result, despite the presence of those fluxes which are naively anticipated to evade the no-go. Thus, our analysis presents a playing ground for constructing/evading the de-Sitter no-go scenarios.

The article is organized as follows: In section \ref{sec_sol-BIs} we present some interesting solutions of the NS-NS Bianchi identities which we further use for deriving the no-go conditions in the upcoming sections. Section \ref{sec_nogo1} presents a type IIA no-go with standard fluxes and its $T$-dual type IIB counterpart which includes non-geometric fluxes as well. In section \ref{sec_nogo2} first we re-derive the fact that one can evade the type IIA no-go-1 with geometric fluxes and Romans mass, and then we $T$-dualize it to study the type IIB counter part. Section \ref{sec_nogo3} presents the relevance of $K3/{\mathbb T}^4$-fibred Calabi Yau threefolds which help in finding a new class of de-Sitter no-go scenarios in both the type II theories. Finally we conclude with the results and observations in section \ref{sec_conclusions}.

\vskip0.2cm
\noindent
{\bf Note:} Let us mention at the outset that we will follow the $T$-dual dictionary from a companion paper \cite{Shukla:2019wfo} which includes the necessary ingredients of the generic formulation of the four-dimensional scalar potentials for the type IIA and the type IIB supergrativities with (non-)geometric fluxes, and this dictionary is placed in the appendix \ref{sec_dictionary}. For the current interests in this article, we will directly utilize the scalar potential for the possible applications in the lights of de-Sitter and inflationary no-go scenarios. Though we attempt to keep the article self-contained, we encourage the interested readers to follow the other relevant details if necessary, e.g. on the superpotential, $D$-terms etc., directly from \cite{Shukla:2019wfo}. 

\section{Solutions of Bianchi identities}
\label{sec_sol-BIs}
In this section we aim to present some interesting solutions of the Bianchi identities satisfied by the various fluxes of the type IIA and IIB theories. The full list of allowed NS-NS fluxes, namely $\{{\rm H}, w, {\rm Q}, {\rm R}\}$ in type IIA and $\{{H}, \omega, {Q}, {R}\}$ in type IIB along with the RR fluxes, namely $\{{\rm F}_0 \equiv m_0, {\rm F}_2 \equiv m^a, {\rm F}_4 \equiv e_a, {\rm F}_6 \equiv e_0\}$ in type IIA and $\{F_0, F_i, F^i, F^0\}$ in type IIB, and their $T$-duality relations are collected in table \ref{tab_fluxTdual}.
\begin{table}[h!]
\begin{center}
\begin{tabular}{|c||c|c|} 
\hline
& & \\
& Type IIA with $D6/O6$  \quad  & \quad Type IIB with $D3/O3$ and $D7/O7$ \\
& &  \\
\hline
\hline
& &  \\
$F$-term & ${\rm H}_0$,  \quad ${\rm H}_k$, \quad ${\rm H}^\lambda$, & $H_0$, \quad $\omega_{a0}$, \quad $\hat{Q}^\alpha{}_0$, \\
fluxes  & & \\
& $w_{a0}$, \quad $w_{ak}$, \quad $w_a{}^\lambda$, & $H_i$, \quad $\omega_{ai}$, \quad $\hat{Q}^\alpha{}_{i}$, \\ 
& & \\
& ${\rm Q}^a{}_0$, \quad ${\rm Q}^a{}_k$, \quad ${\rm Q}^{a \lambda}$, & $H^i$, \quad $\omega_a{}^i$, \quad $\hat{Q}^{\alpha i}$, \\
& & \\
& ${\rm R}_0$,  \quad ${\rm R}_k$, \quad ${\rm R}^\lambda$, & $- H^0$, \quad $- \omega_{a}{}^0$, \quad $- \hat{Q}^{\alpha 0}$, \\
& & \\
& $e_0$,  \quad $e_a$, \quad $m^a$, \quad $m_0$. & $F_0$,  \quad $F_i$, \quad $F^i$, \quad $- F^0$. \\
\hline
& & \\
$D$-term & $\hat{w}_\alpha{}^0$, \quad $\hat{w}_\alpha{}^k$, \quad $\hat{w}_{\alpha \lambda}$, & $-\,R_K$, \quad $-\,Q^a{}_K$, \quad $\hat{\omega}_{\alpha K}$,\\
fluxes & & \\
& $\hat{\rm Q}^{\alpha 0}$, \quad  $\hat{\rm Q}^{\alpha k}$, \quad $\hat{\rm Q}^{\alpha}{}_\lambda$. & $-\,R^K$, \quad $-\,Q^{a K}$, \quad $\hat{\omega}_{\alpha}{}^K$.\\
& &  \\
\hline
& & \\
Complex & \, \, ${\rm N}^0$, \, \,  ${\rm N}^k$, \, \, ${\rm U}_\lambda$, \, \, ${\rm T}^a$. & $S$, \, \, $G^a$, \, \, $T_\alpha$, \, \, $U^i$.\\
Moduli& & \\
\hline
\end{tabular}
\end{center}
\caption{T-duality transformations among the various fluxes and complex variable.}
\label{tab_fluxTdual}
\end{table}
\noindent
Here the flux as well as various moduli are counted via the Hodge numbers as $\alpha \in \{1, 2, .., h^{1,1}_+\}$, $a \in \{1, 2, .., h^{1,1}_-\}$ on both sides, while $\Lambda \in \{0, 1, 2, .., h^{2,1}_-\}$ and $J, K \in  \{1, 2, .., h^{2,1}_+\}$ on the type IIB side, whereas the splitting of the complex structure indices on type IIA side being such that the $k$ and $\lambda$ sum to $h^{2,1}$. The various fluxes appearing in the four-dimensional type IIA supergravity are constrained by the following five classes of NS-NS Bianchi identities \cite{Shukla:2019wfo}, 
\bea
\label{eq:IIABIs2}
& {\bf (I).} \quad & {\rm H}^{\lambda} \, \hat{w}_{\alpha\lambda} = {\rm H}_{\hat{k}} \, \hat{w}_\alpha{}^{\hat{k}}, \\
& {\bf (II).} \quad & {\rm H}^{\lambda} \, \hat{\rm Q}^\alpha{}_{\lambda} = {\rm H}_{\hat{k}} \, \hat{\rm Q}^{\alpha \, \hat{k}}, \qquad w_a{}^\lambda \, \hat{w}_{\alpha \lambda} = w_{a \hat{k}} \, \hat{w}_\alpha{}^{\hat{k}}\,, \nonumber\\
& {\bf (III).} \quad & \hat{\rm Q}^\alpha{}_\lambda \, w_a{}^\lambda = w_{a \hat{k}} \, \hat{\rm Q}^{\alpha \hat{k}}, \qquad {\rm Q}^a{}_{\hat k} \, \hat{w}_\alpha{}^{\hat k} = {\rm Q}^{a \lambda} \, \hat{w}_{\alpha \lambda}, \nonumber\\
& & \hat{w}_{\alpha\lambda}\, \hat{\rm Q}^{\alpha \hat{k}} = \hat{\rm Q}^\alpha{}_\lambda \, \hat{w}_\alpha{}^{\hat k}, \quad \hat{w}_{\alpha \lambda} \, \hat{\rm Q}^\alpha{}_\rho = \hat{\rm Q}^\alpha{}_\lambda \, \hat{w}_{\alpha \rho}, \quad \hat{w}_\alpha{}^{\hat k} \, \hat{\rm Q}^{\alpha \hat{k^\prime}} = \hat{\rm Q}^{\alpha \hat{k}} \, \hat{w}_\alpha{}^{\hat{k^\prime}}, \nonumber\\
& & {\rm R}^\lambda \, {\rm H}_{\hat{k}} - {\rm H}^\lambda \, {\rm R}_{\hat{k}} + w_a{}^\lambda \, {\rm Q}^a{}_{\hat{k}} - {\rm Q}^{a \lambda} \, w_{a \hat{k}} =0, \nonumber\\
& & {\rm H}_{[\hat{k}} \, {\rm R}_{\hat{k^\prime}]} + {\rm Q}^a{}_{[\hat{k}} \, w_{a \hat{k^\prime}]} = 0, \qquad  {\rm H}^{[\lambda} \, {\rm R}^{\rho]} +  {\rm Q}^{a [\lambda} \, w_a{}^{\rho]} = 0, \nonumber\\
& {\bf (IV).} \quad & {\rm R}^\lambda \, \hat{w}_{\alpha \lambda} = {\rm R}_{\hat k} \, \hat{w}_\alpha{}^{\hat k}, \qquad {\rm Q}^{a \lambda} \, \hat{\rm Q}^\alpha{}_\lambda = {\rm Q}^a{}_{\hat k} \, \hat{\rm Q}^{\alpha \hat{k}} \,, \nonumber\\
& {\bf (V).} \quad & {\rm R}^\lambda \, \hat{\rm Q}^\alpha{}_\lambda = {\rm R}_{\hat k} \, \hat{\rm Q}^{\alpha \hat{k}}\,. \nonumber
\eea
Similarly on type IIB side, we have the following five classes of Bianchi identities \cite{Robbins:2007yv}, 
\bea
\label{eq:IIBBIs2}
& {\bf (I).} \quad & H_\Lambda \, \omega_{a}{}^{\Lambda} = H^\Lambda \, \omega_{\Lambda a}, \\
& {\bf (II).} \quad & H^\Lambda \, \hat{Q}_\Lambda{}^\alpha = H_\Lambda \hat{Q}^{\alpha \Lambda}, \qquad \omega_{a}{}^{\Lambda} \, \omega_{b \Lambda} = \omega_{b}{}^{\Lambda} \, \omega_{a \Lambda}, \qquad \hat{\omega}_{\alpha}{}^{K} \, \hat{\omega}_{\beta K} = \hat{\omega}_{\beta}{}^{K} \, \hat{\omega}_{\alpha K}, \nonumber\\
& {\bf (III).} \quad & \omega_{a \Lambda} \, \hat{Q}^{\alpha \Lambda} = \omega_{a}{}^{\Lambda} \, \hat{Q}^\alpha{}_{\Lambda}, \quad Q^{a K} \, \hat{\omega}_{\alpha K} = Q^{a}{}_{K} \, \hat{\omega}_{\alpha}^{K}, \nonumber\\
& & H_\Lambda \, R_K + \omega_{a \Lambda} \, Q^a{}_K + \hat{Q}^\alpha{}_\Lambda \, \hat{\omega}_{\alpha K} = 0, \qquad H^\Lambda \, R_K + \omega_{a}{}^{ \Lambda} \, Q^a{}_K + \hat{Q}^{\alpha{}\Lambda} \, \hat{\omega}_{\alpha K} = 0, \nonumber\\
& & H_\Lambda \, R^K + \omega_{a \Lambda} \, Q^{a{}K} + \hat{Q}^\alpha{}_\Lambda \, \hat{\omega}_{\alpha}{}^{K} = 0, \qquad H^\Lambda \, R^K + \omega_{a}{}^{ \Lambda} \, Q^{a K} + \hat{Q}^{\alpha{}\Lambda} \, \hat{\omega}_{\alpha}{}^{K} = 0, \nonumber\\
& {\bf (IV).} \quad & R^K \, \hat{\omega}_{\alpha K} = R_K \, \hat{\omega}_{\alpha}{}^{K}, \qquad \hat{Q}^{\alpha\Lambda} \, \hat{Q}^\beta{}_{\Lambda} = \hat{Q}^{\beta \Lambda} \, \hat{Q}^\alpha{}_{\Lambda}, \qquad  Q^{a K} \, Q^{b}{}_{K} = Q^{b K} \, Q^{a}{}_{K}, \nonumber\\
& {\bf (V).} \quad & R_K \, Q^{a K} = R^K \, Q^{a}{}_{K}\,. \nonumber
\eea
\noindent
First we argue how by choosing a certain type of involution can project out many flux components and hence can indeed simplify the generic set of identities, for which finding solutions becomes rather easier. Moreover we present another set of solutions which we call as `special solution' for both the type IIA and type IIB theories. They are very peculiar in many aspects as we will elaborate later on.

\subsection{Simple solutions}
The set of type IIA Bianchi identities given in eqn. (\ref{eq:IIABIs2}) suggests that if one choses the anti-holomorphic involution such that the even $(1,1)$-cohomology sector is trivial, which is very often the case one considers for simple phenomenological model \cite{Villadoro:2005cu, Blumenhagen:2013hva, Gao:2017gxk, Gao:2018ayp}, then only the following Bianchi identities remain non-trivial, 
\bea
\label{eq:IIAsimplesol1}
& & {\rm R}^\lambda \, {\rm H}_{\hat{k}} - {\rm H}^\lambda \, {\rm R}_{\hat{k}} + w_a{}^\lambda \, {\rm Q}^a{}_{\hat{k}} - {\rm Q}^{a \lambda} \, w_{a \hat{k}} =0, \\
& & {\rm H}_{[\hat{k}} \, {\rm R}_{\hat{k^\prime}]} + {\rm Q}^a{}_{[\hat{k}} \, w_{a \hat{k^\prime}]} = 0, \qquad  {\rm H}^{[\lambda} \, {\rm R}^{\rho]} +  {\rm Q}^{a [\lambda} \, w_a{}^{\rho]} = 0\,. \nonumber
\eea
In such a situation, there will be no $D$-term contributions generated to the scalar potential as all the fluxes relevant for $D$-terms have $\alpha \in h^{1,1}_+$ indices, and hence are projected out.

For the $T$-dual of the above type IIA setting, one needs to look at the set of type IIB Bianchi identities given in eqn. (\ref{eq:IIBBIs2}) which suggests that if one choses the holomorphic involution such that the even $(2,1)$-cohomology sector is trivial, then only the following Bianchi identities remain non-trivial, 
\bea
\label{eq:IIBsimplesol1}
& & H_\Lambda \, \omega_{a}{}^{\Lambda} = H^\Lambda \, \omega_{\Lambda a}, \qquad H^\Lambda \, \hat{Q}_\Lambda{}^\alpha = H_\Lambda \hat{Q}^{\alpha \Lambda}, \qquad \omega_{a}{}^{\Lambda} \, \omega_{b \Lambda} = \omega_{b}{}^{\Lambda} \, \omega_{a \Lambda}, \\
& & \omega_{a \Lambda} \, \hat{Q}^{\alpha \Lambda} = \omega_{a}{}^{\Lambda} \, \hat{Q}^\alpha{}_{\Lambda}, \qquad \hat{Q}^{\alpha\Lambda} \, \hat{Q}^\beta{}_{\Lambda} = \hat{Q}^{\beta \Lambda} \, \hat{Q}^\alpha{}_{\Lambda}, \nonumber
\eea
which are in a one-to-one correspondence with those in eqn. (\ref{eq:IIAsimplesol1}). In such a situation, there will be no $D$-term generated as all the fluxes with $\{J, K \} \in h^{2,1}_+$ indices are projected out. Moreover, on top of this if the holomorphic involution is chosen to result in a trivial odd $(1,1)$-cohomology, which corresponds to a situation with the absence of odd moduli $G^a$ on the type IIB side and is also very often studied case for being simplistic in nature (e.g. see \cite{Aldazabal:2006up, Blumenhagen:2013hva, Shukla:2016xdy, Betzler:2019kon}), then there are only two Bianchi identities to worry about and they are given as under,
\bea
\label{eq:IIBsimplesol2}
& & H^\Lambda \, \hat{Q}_\Lambda{}^\alpha = H_\Lambda \hat{Q}^{\alpha \Lambda}, \qquad \hat{Q}^{\alpha\Lambda} \, \hat{Q}^\beta{}_{\Lambda} = \hat{Q}^{\beta \Lambda} \, \hat{Q}^\alpha{}_{\Lambda}. 
\eea
This further simplification on type IIB side corresponds to the absence of $N^k$ moduli on the type IIA side, and so is the case for the corresponding fluxes which couple to $N^k$ through the superpotential. This leads to two Bianchi identities on the type IIA side which happen to be $T$-dual to those presented in eqn. (\ref{eq:IIBsimplesol2}), and are given as,
\bea
\label{eq:IIAsimplesol2}
& & \hskip-1cm {\rm R}^\lambda \, {\rm H}_0 - {\rm H}^\lambda \, {\rm R}_0 + w_a{}^\lambda \, {\rm Q}^a{}_0 - {\rm Q}^{a \lambda} \, w_{a 0} = 0, \qquad  {\rm H}^{[\lambda} \, {\rm R}^{\rho]} +  {\rm Q}^{a [\lambda} \, w_a{}^{\rho]} = 0\,. 
\eea
These `simple' solutions of the Bianchi identities based on some specific choice of orientifold involution leads to some interesting scenarios both in type IIA and type IIB theories.

\subsection{IIA  with `special solution' $\equiv$ IIB with geometric-flux $\equiv$ $\exists$ dS no-go}
From the set of type IIA Bianchi identities given in eqn. (\ref{eq:IIABIs2}), one can observe that several Bianchi identities appear in the form of orthogonal symplectic vectors and therefore half of the flux components can be set to zero by performing appropriate symplectic rotations\footnote{See \cite{Ihl:2006pp, Ihl:2007ah, Robbins:2007yv} also, for more arguments in this regard relating to dyonic Black hole charges.}. The same is equivalent to setting some fluxes, say those with upper $h^{2,1}$ indices, to zero as we present below,
\bea
\label{eq:condHalf-IIA}
& & {\rm H}^\lambda = 0, \qquad \hat{w}_\alpha{}^0 = \hat{w}_\alpha{}^k = w_a{}^\lambda = 0, \\
& & {\rm R}^\lambda = 0, \qquad \hat{\rm Q}^{\alpha 0} =\hat{\rm Q}^{\alpha k} = {\rm Q}^{a \lambda} = 0. \nonumber
\eea
This is what we call as `special solution'. Now, using these `special' flux choices in eqn. (\ref{eq:condHalf-IIA}) results in the fact that all the type IIA Bianchi identities except the following three are trivially satisfied,
\bea
\label{eq:}
& & {\rm H}_{[0} \, {\rm R}_{{k}]} + {\rm Q}^a{}_{[0} \, w_{a {k}]} = 0, \\
& & {\rm H}_{[{k}} \, {\rm R}_{{k^\prime}]} + {\rm Q}^a{}_{[{k}} \, w_{a {k^\prime}]} = 0, \nonumber\\
& & \hat{w}_{\alpha \lambda} \, \hat{\rm Q}^\alpha{}_\rho = \hat{\rm Q}^\alpha{}_\lambda \, \hat{w}_{\alpha \rho}\,. \nonumber
\eea
This makes a huge simplification in the generic complicated flux constraints. Now considering the $T$-dual of the type IIA `special' flux choice, as given in eqn. (\ref{eq:condHalf-IIA}), turns out to be equivalent to switching-off the following flux components on the type IIB side,
\bea
\label{eq:Tdual-condHalf-IIA}
& & \hat{Q}^\alpha{}_0 = \hat{Q}^\alpha{}_{i} = Q^a{}_K = 0, \qquad R_K = 0, \\
& & Q^{\alpha i} = \hat{Q}^{\alpha 0} = Q^{a K} = 0, \qquad R^K = 0, \nonumber
\eea
which means setting all the non-geometric ($Q$ as well as $R$) fluxes to zero on the type IIB side. Moreover, using the $T$-dual flux choice on type IIB side as given in eqn. (\ref{eq:Tdual-condHalf-IIA}), one finds that the set of Bianchi identities on the type IIB side are reduced into the following three constraints,
\bea
\label{eq:}
& & \hskip-1cm H_\Lambda \, \omega_{a}{}^{\Lambda} = H^\Lambda \, \omega_{\Lambda a}, \qquad \omega_{a}{}^{\Lambda} \, \omega_{b \Lambda} = \omega_{b}{}^{\Lambda} \, \omega_{a \Lambda}, \qquad \hat{\omega}_{\alpha}{}^{K} \, \hat{\omega}_{\beta K} = \hat{\omega}_{\beta}{}^{K} \, \hat{\omega}_{\alpha K}\,,
\eea
which is very much expected as there are no non-zero ${\rm Q}$ and ${\rm R}$ flux components present in the current setting. As a side remark, let us point out that if the involutions are considered as per the choices earlier explained as `simple solutions', i.e. those without $D$-terms, then there remains just two identities on the two sides,
\bea
\label{eq:}
& {\rm \bf IIA:} & \qquad {\rm H}_{[0} \, {\rm R}_{{k}]} + {\rm Q}^a{}_{[0} \, w_{a {k}]} = 0, \qquad {\rm H}_{[{k}} \, {\rm R}_{{k^\prime}]} + {\rm Q}^a{}_{[{k}} \, w_{a {k^\prime}]} = 0, \\
& {\rm \bf IIB:} & \qquad H_\Lambda \, \omega_{a}{}^{\Lambda} = H^\Lambda \, \omega_{\Lambda a}, \qquad \qquad \,\,\omega_{a}{}^{\Lambda} \, \omega_{b \Lambda} = \omega_{b}{}^{\Lambda} \, \omega_{a \Lambda}\, \nonumber
\eea
and even the above ones are absent if one sets $a =0$, i.e. no $G^a$ moduli in IIB and equivalently no ${\rm N}^k$ moduli in IIA. Thus with some orientifold setting one can have `special solutions' in which all the Bianchi identities are trivial ! Note that all these identities are well in line with the $T$-duality transformations inherited from their generic structure before taking any simplification.

\subsection*{A no-go condition for de-Sitter and slow-roll inflation:}
As we have seen that the type IIA non-geometric setup with `special solution' leads to a type IIB setup without any non-geometric flux. Now, following from the table \ref{tab_scalar-potential} of the dictionary \ref{sec_dictionary}, the type IIB scalar potential can be expressed as a sum of the following pieces,
\bea
\label{eq:no-goIIB}
& & V_{\rm IIB}^{\rm RR} = \frac{e^{4\phi}}{4\,{\cal V}^2\, {\cal U}}\Bigl[f_0^2 + {\cal U}\, f^i \, {\cal G}_{ij} \, f^j + {\cal U}\, f_i \, {\cal G}^{ij} \,f_j + {\cal U}^2\, (f^0)^2\Bigr],\\
& & V_{\rm IIB}^{\rm NS1} = \frac{e^{2\phi}}{4\,{\cal V}^2\,{\cal U}}\Bigl[h_0^2 + {\cal U}\, h^i \, {\cal G}_{ij} \, h^j + {\cal U}\, h_i \, {\cal G}^{ij} \,h_j + {\cal U}^2\, (h^0)^2 \Bigr], \nonumber\\
& & V_{\rm IIB}^{\rm NS2} = \frac{e^{2\phi}}{4\,{\cal V}^2\,{\cal U}}\Bigl[\, {\cal V}\, {\cal G}^{ab}\,(h_{a0} \, h_{b0} + \frac{l_i\, l_j}{4} \,h_a{}^i\, h_b{}^j  + \, h_{ai} \, h_{bj} \, u^{i}\, u^{j} + {\cal U}^2\, h_a{}^0\, h_b{}^0 \nonumber\\
& & \quad \qquad - \, \frac{l_i}{2}\, h_a{}^i\, h_{b0} - \frac{l_i}{2} \,h_{a0}\, h_b{}^i - {\cal U} \, u^i \, h_a{}^0 \, h_{bi} - {\cal U} \, u^i \, h_b{}^0 \, h_{ai}) \Bigr], \nonumber\\
& & V_{\rm IIB}^{\rm loc} = \frac{e^{3\phi}}{2\, {\cal V}^2} \left[f^0 h_0 - f^i h_i + f_i h^i - f_0 h^0\right], \nonumber\\
& & V_{\rm IIB}^{D} = \frac{e^{2\phi}}{4\,{\cal V}^2} \Bigl[{t}^\alpha \, {t}^\beta \, ( \hat{h}_{\alpha J} \,{\cal G}^{JK} \, \hat{h}_{\beta K} +  \, \hat{h}_\alpha{}^J \,{\cal G}_{JK} \, \hat{h}_{\beta}{}^K) \Bigr], \nonumber
\eea
where $f_0, \, f_i,\, f^i,\, f^0,\, h_0, \, h_i,\, h^i,\, h^0,\, h_{a0},\, h_{ai},\, h_a{}^0,\, h_a{}^i, \, \hat{h}_{\alpha K}$ and $\hat{h}_\alpha{}^K$ are the axionic flux orbits as defined in table \ref{tab_IIB-Fluxorbits}. However as they do not depend on any of the saxions, it is not relevant to give their explicit lengthy details. Also note that in this orientifold we have the following axionic flux orbits of table table \ref{tab_IIB-Fluxorbits} being identically zero on the type IIB side,
\bea
& & \hskip-1cm h^\alpha{}_0 = h^\alpha{}_i = h^{\alpha i} = h^{\alpha0} = 0, \quad \quad \quad h_K{}^0 = h^{K0} = 0\,.
\eea
For studying the scalar potential in eqn. (\ref{eq:no-goIIB}), let us extract the volume factor by introducing a new modulus $\rho$ via defining the two-cycle volume moduli as $t^\alpha = \rho\, \gamma^\alpha$ where $\gamma^\alpha$ is angular K\"ahler moduli satisfying the constraint $\ell_{\alpha\beta\gamma} \gamma^\alpha \gamma^\beta \gamma^\gamma = 6$. This leads to the overall volume being given as ${\cal V} = \rho^3$ and the volume dependent moduli space metric being simplified as,
\bea
& & {\cal G}^{ab} = - \hat\ell^{ab}= - \frac{1}{\rho} (\hat\ell_{\alpha ab} \gamma^\alpha)^{-1}.
\eea
Also note that the moduli space metric ${\cal G}^{JK}$ and its inverse ${\cal G}_{JK}$ are independent of any of the volume moduli, and in particular on the $\rho$ modulus as well. Subsequently the scalar potential can be expressed as under,
\bea
& & V = V_1 + V_2 + V_3 + V_4\,,
\eea
where defining a new variable $\tau = e^{-\phi} \sqrt{\cal V} = e^{-\phi} \rho^{3/2}$, the above four pieces are given as,
\bea
& & V_1 = \frac{A_1}{\tau^4}, \qquad V_2 = \frac{A_2}{\tau^2 \, \rho^3}, \qquad V_3 =  \frac{A_3}{\tau^2 \, \rho}, \qquad V_4= \frac{A_4}{\tau^3\, \rho^{3/2}}\,.
\eea
Here $A_i$'s depend on the complex structure moduli and the angular K\"ahler moduli but not on any of the $\tau$ and $\rho$ moduli. In addition one has $A_1 \geq 0, \, A_2 \geq 0$ however signs of $A_3$ and $A_4$ are not fixed. Also note that we have combined the two pieces $V_{\rm IIB}^{\rm NS2}$ and $V_{\rm IIB}^{D}$ as they have the same scaling for the $\rho$ and $\tau$ moduli. This leads to the following relation,
\bea
& & -3\, \tau\, \partial_\tau V - \rho \, \partial_\rho V = 12 V_1 + 9 V_2 + 7 V_3 + \frac{21}{2}\, V_4.
\eea
This apparently shows that the necessary condition for the de-Sitter no-go scenario, which one usually gets in the $(\tau, \rho)$-plane, is evaded. But after checking trace and determinants of the Hessian in the $(\tau, \rho)$-plane, one finds that determinant of the Hessian evaluated at the extremum is never positive, and hence confirming a no-go case due to the presence of tachyons. Such a type IIB setup with $D3/D7$ and $O3/O7$ having $F_3, H_3$ and the geometric flux has been also studied in \cite{Shiu:2011zt, Garg:2018reu}, where it was concluded that no stable de-Sitter vacua can be realized in this type IIB setting. Thus from our $T$-duality rules, we conclude the following de-Sitter no-go condition on the dual type IIA side:
\begin{mdframed}
\noindent
{\bf Type IIA No-Go theorem:}  In the framework of non-geometric type IIA orientifold compactification with $O6$ planes, one cannot have a de-Sitter solution by merely considering the RR flux $F_0, F_2, F_4, F_6$ along with the `special solutions' of the NS-NS Bianchi identities.
\end{mdframed} 
Note that given the fact that there are certain non-geometric flux components present in the dual type IIA side despite corresponding to the special solutions of the Bianchi identities, this de-Sitter no-go condition would not have been possible to guess a priory the explicit computations are done, but from the type IIB side it is not hard to invoke. 

\subsection{IIB with `special solution' $\equiv$ IIA with geometric-flux $\equiv$ $\nexists$ dS no-go}
Similar to the type IIA case, one can observe from the eqn. (\ref{eq:IIBBIs2}) that many of the type IIB Bianchi identities also appear in the form of orthogonal symplectic vectors and therefore half of the flux components can be rotated away, as presented below:
\bea
\label{eq:condHalf-IIB}
& & H^0 = 0 = H^i, \qquad \omega_a{}^0 = 0 = \omega_a{}^i, \qquad \hat{Q}^{\alpha 0} = 0 = \hat{Q}^{\alpha i}, \\
& & \hat{\omega}_{\alpha}{}^K =0, \qquad Q^{a K} =0, \qquad R^K = 0\,. \nonumber
\eea
Now, one can observe that using the `special' flux choice in eqn. (\ref{eq:condHalf-IIB}) results in the fact that all the type IIB Bianchi identities except the following two are trivially satisfied,
\bea
\label{eq:condHalf-IIB2}
& & H_0 \, R_K + \omega_{a 0} \, Q^a{}_K + \hat{Q}^\alpha{}_0 \, \hat{\omega}_{\alpha K} = 0, \\
& & H_i \, R_K + \omega_{a i} \, Q^a{}_K + \hat{Q}^\alpha{}_i \, \hat{\omega}_{\alpha K} = 0\,. \nonumber
\eea
Moreover, the type IIB `special solution' as given in eqn. (\ref{eq:condHalf-IIB}) is equivalent to switching-off the following $T$-dual fluxes on the type IIA side,
\bea
\label{eq:Tdual-condHalf-IIB}
& & {\rm Q}^a{}_0 = {\rm Q}^a{}_k = {\rm Q}^{a \lambda} = 0, \qquad  \hat{\rm Q}^{\alpha 0} = \hat{\rm Q}^{\alpha k} = \hat{\rm Q}^{\alpha}{}_\lambda =0, \\
& & {\rm R}_0 = {\rm R}_k = {\rm R}^\lambda =0\,. \nonumber
\eea
This immediately implies that type IIB `special solutions' correspond to setting all the non-geometric fluxes to zero on the type IIA side.  Further, using the $T$-duality on type IIB side, the two constraints given in eqn. (\ref{eq:condHalf-IIB2}) translates into the following two constraints on the type IIA side,
\bea
\label{eq:Tdual-condHalf-IIB2}
& & {\rm H}^{\lambda} \, \hat{w}_{\alpha\lambda} = {\rm H}_{\hat{k}} \, \hat{w}_\alpha{}^{\hat{k}}, \qquad w_a{}^\lambda \, \hat{w}_{\alpha \lambda} = w_{a \hat{k}} \, \hat{w}_\alpha{}^{\hat{k}}\,,
\eea
which is very much expected as there are no non-zero ${\rm Q}$- and ${\rm R}$-flux components present in this setting. As a side remark, one can observe that for a trivial even $(2,1)$-cohomology on type IIB side, `special solution' is sufficient to satisfy all the flux constraints as the constraints in eqn. (\ref{eq:condHalf-IIB2}) get trivial. On the $T$-dual type IIA side, this would mean to have the even-$(1,1)$-cohomology trivial and so trivially satisfying the eqn. (\ref{eq:Tdual-condHalf-IIB2}).  
A summary of the results of this section has been presented in table \ref{tab_special-flux-sol}.
\begin{table}[h!]
\begin{center}
\begin{tabular}{|c||c||c|c|} 
\hline
& & & \\
Scenario & $\exists$ no-go & Type IIA with $D6/O6$  \quad  & \quad Type IIB with  \\
& & & $D3/O3$ and $D7/O7$ \\
\hline
\hline
& & & \\
Type IIA & Yes & ${\rm H}_0$,  \quad $w_{a0}$, \quad ${\rm Q}^a{}_0$, \quad ${\rm R}_0$,  & $H_0$, \quad $H_i$, \quad $H^i$, \quad $- H^0$,   \\
with & & & \\
special & & ${\rm H}_k$, \quad $w_{ak}$, \quad ${\rm Q}^a{}_k$, \quad ${\rm R}_k$, & $\omega_{a0}$, \quad $\omega_{ai}$, \quad $\omega_a{}^i$, \quad $- \omega_{a}{}^0$, \\
solutions & & & \\
& & $e_0$,  \quad $e_a$, \quad $m^a$, \quad $m_0$. & $F_0$,  \quad $F_i$, \quad $F^i$, \quad $- F^0$. \\
& & & \\
& & $\hat{w}_{\alpha \lambda}$, \quad $\hat{\rm Q}^{\alpha}{}_\lambda$. & $\hat{\omega}_{\alpha K}$, \quad $\hat{\omega}_{\alpha}{}^K$.\\
& & & \\
& & & (Type IIB with geometric flux) \\
& & & \\
\hline
\hline
& & & \\
Type IIB & No & ${\rm H}_0$,  \quad ${\rm H}_k$, \quad ${\rm H}^\lambda$, & $H_0$, \quad $\omega_{a0}$, \quad $\hat{Q}^\alpha{}_0$, \\
with & & & \\
special & & $w_{a0}$, \quad $w_{ak}$, \quad $w_a{}^\lambda$, & $H_i$, \quad $\omega_{ai}$, \quad $\hat{Q}^\alpha{}_{i}$, \\ 
solution & & & \\
& & $e_0$,  \quad $e_a$, \quad $m^a$, \quad $m_0$. & $F_0$,  \quad $F_i$, \quad $F^i$, \quad $- F^0$. \\
& & & \\
& & $\hat{w}_\alpha{}^0$, \quad $\hat{w}_\alpha{}^k$, \quad $\hat{w}_{\alpha \lambda}$, & $-\,R_K$, \quad $-\,Q^a{}_K$, \quad $\hat{\omega}_{\alpha K}$,\\
& & & \\
& & (Type IIA with geometric flux) & \\
\hline
\end{tabular}
\end{center}
\caption{Possible non-zero fluxes in the special solutions of Bianchi identities.}
\label{tab_special-flux-sol}
\end{table}

\noindent

\section{No-Go 1}
\label{sec_nogo1}
In this section we present the de-Sitter no-go scenario realized in the context of type IIA flux compactification with the inclusion of the NS-NS $H_3$ flux, and the standard R-R fluxes, namely the $F_0, F_2, F_4, F_6$ flux \cite{Hertzberg:2007wc}. First we revisit the ingredients of the no-go condition and then we will $T$-dualize the same to investigate the no-go condition in the type IIB theory.

\subsection{Type IIA with RR-flux and $H_3$-flux}
In the absence of any geometric and non-geometric fluxes in the type IIA flux compactifications, the generic four-dimensional scalar potential presented in the table \ref{tab_scalar-potential} simplifies to a form given as under,
\bea
\label{eq:nogo1-IIA1}
& & \hskip-0.5cm V_{\rm IIA} = \frac{e^{4D}}{4\, {\cal V}}\biggl[{\rm f}_0^2 + {\cal V}\, {\rm f}^a \, \tilde{\cal G}_{ab} \, {\rm f}^b + {\cal V}\, {\rm f}_a \, \tilde{\cal G}^{ab} \,{\rm f}_b + {\cal V}^2\, ({\rm f}^0)^2\biggr]\,\\
& & \hskip0.5cm + \, \frac{e^{2D}}{4\,{\cal V}}\biggl[\frac{{\rm h}_0^2}{\cal U} +\, \tilde{\cal G}^{ij}\,{\rm h}_{i0} \, {\rm h}_{j0} + \,\tilde{\cal G}_{\lambda \rho} {\rm h}^\lambda{}_0 \, {\rm h}^\rho{}_0 \biggr]\, + \frac{e^{3D}}{2\, \sqrt{\cal U}} \left[{\rm f}^0 \, {\rm h}_0 - \frac{k_\lambda}{2} \, {\rm f}^0\, {\rm h}^\lambda{}_0 \right], \nonumber
\eea
where the various ``axionic flux orbits" defined in table \ref{tab_IIA-Fluxorbits} are simplified to the following form,
\bea
& & \hskip-1cm {\rm f}_0  = e_0 + \, {\rm b}^a\, e_a + \frac{1}{2} \, \kappa_{abc} \, {\rm b}^a\, {\rm b}^b \,m^c + \frac{1}{6}\, \kappa_{abc}\,  {\rm b}^a \, {\rm b}^b\, {\rm b}^c \, m_0 - \, \xi^{0} \, {\rm H}_0 - \, \xi^k \, {\rm H}_k - {\xi}_\lambda \, {\rm H}^\lambda \,, \nonumber\\
& & \hskip-1cm {\rm f}_a = e_a + \, \kappa_{abc} \,  {\rm b}^b \,m^c + \frac{1}{2}\, \kappa_{abc}\,  {\rm b}^b\, {\rm b}^c \, m_0\,, \quad {\rm f}^a = m^a + m_0\,  {\rm b}^a\,, \quad  {\rm f}^0 = m_0\,, \\
& & \hskip-1cm {\rm h}_0 = {\rm H}_0 + {\rmz^k} \,{\rm H}_k + \, \frac{1}{2} \, \hat{k}_{\lambda mn} \rmz^m \rmz^n \, {\rm H}^\lambda, \quad {\rm h}_{k0} = {\rm H}_k +  \, \hat{k}_{\lambda k n}\, {\rmz^n} \, {\rm H}^\lambda, \quad {\rm h}^\lambda{}_0 = {\rm H}^\lambda\,. \nonumber
\eea
We further introduce a new modulus $\rho$ through a redefinition in the overall volume (${\cal V}$) of the Calabi Yau threefold by considering the two-cycle volume moduli $t^a$ via $t^a = \rho \, \gamma^a$, where $\gamma^a$'s denote the angular K\"ahler moduli satisfying the constraint $\kappa_{abc}\gamma^a\gamma^b\gamma^c = 6$ implying ${\cal V} = \rho^3$. Now we can extract the volume factor $\rho$ from the K\"ahler moduli space metric and its inverse in the following way, 
\bea
\label{eq:IIAmetric-rho}
& & \hskip-1.5cm \tilde{\cal G}_{ab} = \frac{\kappa_a\, \kappa_b - 4\, {\cal V}\, \kappa_{ab}}{4\,{\cal V}} = \rho \, \tilde{g}_{ab}, \qquad \, \, \tilde{\cal G}^{ab} =  \frac{2\, {\rm t}^a \, {\rm t}^b - 4\, {\cal V}\, \kappa^{ab}}{4\,{\cal V}} = \frac{1}{\rho} \, \tilde{g}^{ab}\,,
\end{eqnarray}
where $\tilde{g}_{ab}$ and the inverse $\tilde{g}^{ab}$ do not depend on $\rho$ modulus. Subsequently the scalar potential in eqn. (\ref{eq:nogo1-IIA1}) can be written as under,
\bea
\label{eq:nogo1-IIA2}
& & V_{\rm IIA} = \frac{e^{4D}}{4\,\rho^3}\biggl[{\rm f}_0^2 + \rho^2\, {\rm f}_a \, \tilde{g}^{ab} \,{\rm f}_b + \rho^4\, {\rm f}^a \, \tilde{g}_{ab} \, {\rm f}^b + \rho^6\, ({\rm f}^0)^2\biggr]\,\\
& & \hskip1cm + \, \frac{e^{2D}}{4\,\rho^3}\biggl[\frac{{\rm h}_0^2}{\cal U} +\, \tilde{\cal G}^{ij}\,{\rm h}_{i0} \, {\rm h}_{j0} + \,\tilde{\cal G}_{\lambda \rho} {\rm h}^\lambda{}_0 \, {\rm h}^\rho{}_0 \biggr]\,+ \frac{e^{3D}}{2\, \sqrt{\cal U}} \left[{\rm f}^0 \, {\rm h}_0 - \frac{k_\lambda}{2} \, {\rm f}^0\, {\rm h}^\lambda{}_0 \right]. \nonumber
\eea
Now for the above potential, one can easily show that the following inequality holds,
\bea
& & \hskip-1cm 3\, \partial_D\,V_{\rm IIA} - \rho \, \partial_\rho V_{\rm IIA} = 9 \, V_{\rm IIA} + \frac{e^{4D}}{4\,\rho^3}\biggl[6\, {\rm f}_0^2 + 4\,  \rho^2\, {\rm f}_a \, \tilde{g}^{ab} \,{\rm f}_b + 2\, \rho^4\, {\rm f}^a \, \tilde{g}_{ab} \, {\rm f}^b \biggr] \geq 9 \, V_{\rm IIA}\,,
\eea
where in the last step we have used the fact that all the additional terms in the bracket are guaranteed to be non-negative. This immediately leads to a de-Sitter no-go theorem because at this extremum $\partial_D\,V_{\rm IIA} = 0 = \partial_\rho V_{\rm IIA}$, the potential is evaluated to take non-positive values as we see below,
\bea
& & V_{\rm IIA}^{\rm ext}  = - \frac{1}{9} \times \frac{e^{4D}}{4\,\rho^3}\biggl[6\, {\rm f}_0^2 + 4\,  \rho^2\, {\rm f}_a \, \tilde{g}^{ab} \,{\rm f}_b + 2\, \rho^4\, {\rm f}^a \, \tilde{g}_{ab} \, {\rm f}^b \biggr] \leq 0\,.
\eea
Moreover, one has the following inequality on the inflationary slow-roll $\epsilon$ parameter,
\bea
& & \epsilon \geq V_{\rm IIA}^{-2} \biggl[\frac{\rho^2}{3} {(\partial_\rho V_{\rm IIA})}^2 + \frac{1}{4} {(\partial_D V_{\rm IIA})}^2 \biggr] \\
& & \hskip0.3cm = V_{\rm IIA}^{-2} \biggl[\frac{1}{39} (3\, \partial_D\,V_{\rm IIA} - \rho \, \partial_\rho V_{\rm IIA})^2 + \frac{1}{52} (\partial_D\,V_{\rm IIA} + 4\, \rho \, \partial_\rho V_{\rm IIA})^2 \biggr] \geq \frac{27}{13}\,. \nonumber
\eea
This clearly forbids the slow-roll inflation in this simplistic framework as proposed in \cite{Hertzberg:2007wc, Flauger:2008ad}.

\subsection{$T$-dual de-Sitter no-go-1 in type IIB}
Now we invoke the $T$-dual of this type IIA no-go scenario and investigate the type IIB side. The type IIB fluxes which are $T$-dual to the non-zero type IIA fluxes are given in table \ref{tab_no-go1}.
\noindent
\begin{table}[H]
\begin{center}
\begin{tabular}{|c||c|c|c|c||c|c|c|} 
\hline
& &&&&&&\\
IIA & $e_0 $ & $e_a$ & $m^a$ & $m_0$ & ${\rm H}_0$ & ${\rm H}_k$ & ${\rm H}^\lambda$ \\
& &&&&&&\\
\hline
& &&&&&&\\
IIB & ${F}_0 $ & ${F}_i$ & ${F}^i$ & $- {F}^0$ & ${H}_0$ & ${\omega}_{a0}$ & $\hat{Q}^\alpha{}_0$ \\
& &&&&&&\\
\hline
\end{tabular}
\end{center}
\caption{Non-zero Type IIA fluxes and their respective $T$-duals for {\bf No-Go 1}.}
\label{tab_no-go1}
\end{table}
\noindent
This shows that type IIB side can generically have all the components of the $F_3$ flux while for the NS-NS sector, there are only the `rigid' fluxes which are allowed, though due to a mixing through the $T$-duality, there are some (non-)geometric flux components present unlike the type IIA case. We call $H_0, \, \omega_{a0}$ and $\hat{Q}^\alpha{}_0$ as `rigid fluxes' because they are the ones which are allowed in a type IIB framework without the complex structure moduli. However by saying this we do not mean that our $T$-dual approach is valid for the rigid Calabi Yau compactification as it is well known that mirror of a  rigid Calabi Yau is not a Calabi Yau \cite{Candelas:1993nd, Sethi:1994ch, Hori:2000kt}. We have studied the scalar potentials arising in rigid compactifications separately in \cite{Shukla:2019akv}, and throughout this work we assume that the compactifications are on non-rigid threefolds. For the present case, this type IIB scenario only reflects the fact that we have just rigid fluxes turned-on setting others to zero, and for this, a no-go should exist. 

Having no (non-)geometric fluxes present, there are no Bianchi identities to satisfy in the type IIA side, and the same is true for the type IIB side as well, despite the presence of some rigid (non-)geometric fluxes\footnote{This is something one would expect from the set of Bianchi identities known to us in the cohomology formulation, though there are several observations based on toroidal examples that there may be a few of the missing identities in this approach \cite{Ihl:2007ah, Robbins:2007yv, Shukla:2016xdy, Gao:2018ayp, Shukla:2019akv}.}. The dual scalar potential for the type IIB side can be read-off from the table \ref{tab_scalar-potential} as under,
\bea
\label{eq:nogo1-IIB1}
& & \hskip-0.5cm V_{\rm IIB} = \frac{e^{4\phi}}{4\,{\cal V}^2\, {\cal U}}\biggl[f_0^2 + {\cal U}\, f_i \, {\cal G}^{ij} \,f_j + {\cal U}\, f^i \, {\cal G}_{ij} \, f^j + {\cal U}^2\, (f^0)^2\biggr]\,\\
& & \hskip0.5cm + \frac{e^{2\phi}}{4\,{\cal V}^2\,{\cal U}}\biggl[h_0^2  + \, {\cal V}\, {\cal G}^{ab}\,h_{a0} \, h_{b0} + \, {\cal V} \,{\cal G}_{\alpha \beta}\,h^\alpha{}_0 \, h^\beta{}_0 \biggr]\, + \frac{e^{3\phi}}{2\,{\cal V}^2} \left[{f}^0 \, {h}_0 - \frac{\ell_\alpha}{2} \, {f}^0\, {h}^\alpha{}_0 \right]\,, \nonumber
\eea
where the simplified axionic flux orbits following from table \ref{tab_IIB-Fluxorbits} are given as under,
\bea
& & \hskip-0.5cm f_0 = F_0 + v^i\, {F}_i + \frac{1}{2}\, l_{ijk}\, v^j\, v^k \, {F}^i\, - \frac{1}{6}\, l_{ijk}\, v^i \, v^j\, v^k  \, {F}^0 \\
& & \hskip0.5cm - \omega_{a0} \, {c}^a - \hat{Q}^\alpha{}_0 \, \hat{c}_\alpha   - \, c_0 \, \Big(H_0 + \omega_{a0} \, {b}^a + \frac{1}{2}\, \hat{\ell}_{\alpha a b}\, b^a b^b \, \hat{Q}^\alpha{}_0 \Bigr)\,, \nonumber\\
& & \hskip-0.5cm f_i = {F}_i +\, l_{ijk}\, v^j \, {F}^k - \frac{1}{2}\, l_{ijk}\, v^j\, v^k \, {F}^0 , \quad f^i = F^i - v^i\, F^0, \quad  f^0 = -\, F^0\,, \nonumber\\
& & \hskip-0.5cm h_0 = H_0 + \omega_{a0} \, {b}^a + \frac{1}{2}\, \hat{\ell}_{\alpha a b}\, b^a b^b \, \hat{Q}^\alpha{}_0, \qquad h_{a0} = \omega_{a0} + \hat{Q}^\alpha{}_0 \, \hat{\ell}_{\alpha a b}\, b^b, \qquad h^\alpha{}_0 = \hat{Q}^\alpha{}_0\,. \nonumber
\eea
Although the no-scale structure on type IIB is broken by the presence of the non-zero $\hat{Q}^\alpha{}_0$-flux which couples to $T_\alpha$ moduli in the superpotential and subsequently been reflected via the appearance of the moduli space metric ${\cal G}_{\alpha\beta}$ in the scalar potential (\ref{eq:nogo1-IIB1}), but that would not lead to a de-Sitter solution as suggested by the dual type IIA side. Thus the type IIA no-go condition tells us something interesting and harder to guess a priory on the type IIB side. 

In order to check that this duality based claim is true, all we need to do is to swap the role of the K\"ahler moduli with complex structure moduli. On that line, similar to the case of volume modulus ${\cal V}$, now we defined a new modulus $\sigma$ from the saxion of the complex structure moduli such that $u^i = \sigma\, \lambda^i$, which leads to ${\cal U} = \sigma^3$ subject to a condition: $l_{ijk}\, \lambda^i \, \lambda^j \, \lambda^k =6$ satisfied by the angular complex structure moduli $\gamma^i$ on type IIB side. Now we can extract the $\sigma$ factor from the complex structure moduli space metric and its inverse in the following way, 
\bea
\label{eq:IIBmetric-sigma}
& & \hskip-1.5cm {\cal G}_{ij} = \frac{l_i\, l_j - 4\, {\cal U}\, l_{ij}}{4\,{\cal U}} = \sigma \, {g}_{ij}, \qquad \, \, {\cal G}^{ij} =  \frac{2\, {u}^i \, {u}^j - 4\, {\cal U}\, l^{il}}{4\,{\cal U}} = \frac{1}{\sigma} \, {g}^{ij}\,,
\eea
where $g_{ij}$ and $g^{ij}$ depends only on the angular complex structure moduli and not on the $\sigma$ modulus. Using this information the scalar potential in eqn. (\ref{eq:nogo1-IIB1}) can be written as under,
\bea
& & \hskip-1cm V_{\rm IIB} = \frac{e^{4\phi}}{4\,{\cal V}^2\, \sigma^3}\biggl[f_0^2 + \sigma^2\, f_i \, {g}^{ij} \,f_j + \sigma^4\, f^i \, {g}_{ij} \, f^j + \sigma^6\, (f^0)^2\biggr]\,\\
& & \hskip0.5cm + \frac{e^{2\phi}}{4\,{\cal V}^2\,\sigma^3}\biggl[h_0^2  + \, {\cal V}\, {\cal G}^{ab}\,h_{a0} \, h_{b0} + \, {\cal V} \,{\cal G}_{\alpha \beta}\,h^\alpha{}_0 \, h^\beta{}_0 \biggr]\, + \frac{e^{3\phi}}{2\,{\cal V}^2} \left[{f}^0 \, {h}_0 - \frac{\ell_\alpha}{2} \, {f}^0\, {h}^\alpha{}_0 \right]. \nonumber
\eea
Subsequently it is not hard to show that the following inequality holds,
\bea
& & \hskip-1.5cm 3\, \partial_\phi \,V_{\rm IIB} - \sigma \, \partial_\sigma V_{\rm IIB} = 9 V_{\rm IIB} + \frac{e^{4\phi}}{4\,{\cal V}^2\, \sigma^3}\biggl[f_0^2 + \, \sigma^2\, f_i \, {g}^{ij} \,f_j + \, \sigma^4\, f^i \, {g}_{ij} \, f^j  \biggr] \geq 9 V_{\rm IIB}\,,
\eea
where in the last step we have used the fact that all the additional terms in bracket are guaranteed to be positive semidefinite. This immediately leads to a de-Sitter no-go theorem as at this extremum $\partial_\phi V_{\rm IIB} = 0 = \partial_\sigma V_{\rm IIB}$, the potential  can only take non-positive values as we see below,
\bea
& & V_{\rm IIB}^{\rm ext}  = - \frac{e^{4\phi}}{2\,{\cal V}^2\, \sigma^3}\biggl[6\, f_0^2 + 4\, \sigma^2\, f_i \, {g}^{ij} \,f_j + 2\, \sigma^4\, f^i \, {g}_{ij} \, f^j  \biggr]  \leq 0\,.
\eea
Thus we are able to prove an interesting de-Sitter no-go theorem on the type IIB side. 
\begin{mdframed}
\noindent
{\bf Type IIB No-Go theorem 1:}  In the framework of type IIB non-geometric flux compactification with $O3/O7$ orientifold planes, one cannot have a de-Sitter solution by considering the RR flux $F_3$ along with the rigid NS-NS flux components $H_0, \,\omega_{a0}$ and $\hat{Q}^\alpha{}_0$ only.
\end{mdframed}

\section{No-Go 2}
\label{sec_nogo2}
In this section we consider another no-go condition found in the type IIA framework, which in addition to the ingredients of the no-go-1 scenario, also includes the geometric flux \cite{Haque:2008jz, Caviezel:2008tf, Flauger:2008ad}, and subsequently we will $T$-dualize the same to invoke its type IIB counterpart. 

\subsection{Type IIA with RR-flux, $H_3$-flux and $\omega$-flux}
This type IIA de-Sitter no-go scenario includes the NS-NS $H_3$ flux, geometric flux $w$, and the standard R-R fluxes, namely the $F_0, \, F_2, \, F_4$ and the $F_6$ flux \cite{Haque:2008jz, Caviezel:2008tf, Flauger:2008ad}. However, there are no non-geometric fluxes turned-on, i.e. ${\rm Q}^{a}{}_{\hat{k}} = {\rm Q}^{a\lambda} = \hat{{\rm Q}}^{\alpha \hat{k}} = \hat{{\rm Q}}^\alpha{}_\lambda = 0$ and ${\rm R}_{\hat{k}} = 0 = {\rm R}^\lambda$. In order to get the scalar potential from our generic formula in table \ref{tab_scalar-potential} one has to simply set the following flux orbits to zero,
\bea
\label{eq:axionic-flux-nogo21}
& & \hskip-1cm {\rm h}^a = 0 = {\rm h}^0, \qquad  {\rm h}^a{}_k = 0 = {\rm h}_k{}^0, \qquad  {\rm h}^{a\lambda} = 0 =  {\rm h}^{\lambda 0}\,, \qquad \hat{{\rm h}}^{\alpha0} = 0 = \hat{{\rm h}}^\alpha{}_\lambda,
\eea
where the last two fluxes are parts of the $D$-term contributions via the ${\rm Q}$ flux. Setting off these non-geometric fluxes in eqn. (\ref{eq:axionic-flux-nogo21}), the generic scalar potential given in table \ref{tab_scalar-potential} can be simplified to take a form given as under,
\bea
\label{eq:main4IIA-nogo2gen}
& & \hskip-0.3cm V_{\rm IIA} = \frac{e^{4D}}{4\,{\cal V}}\biggl[{\rm f}_0^2 + {\cal V}\, {\rm f}_a \, \tilde{\cal G}^{ab} \,{\rm f}_b + {\cal V}\, {\rm f}^a \, \tilde{\cal G}_{ab} \, {\rm f}^b + {\cal V}^2\, ({\rm f}^0)^2\biggr] + \frac{e^{2D}}{4\,{\cal V}}\biggl[\frac{{\rm h}_0^2}{\cal U} +\, \tilde{\cal G}^{ij}\,{\rm h}_{i0} \, {\rm h}_{j0} + \,\tilde{\cal G}_{\lambda \rho} {\rm h}^\lambda{}_0 \, {\rm h}^\rho{}_0 \biggr]\,\nonumber\\
& & \quad \quad + \, \frac{e^{2D}}{4\,{\cal V}\,}\biggl[{\rm t}^{a}\, {\rm t}^{b} \left(\frac{{\rm h}_a \, {\rm h}_b}{\cal U} + \, \tilde{\cal G}^{ij}\, {\rm h}_{ai} \, {\rm h}_{bj} \, + \,\tilde{\cal G}_{\lambda \rho}\, {\rm h}_a{}^\lambda\, {\rm h}_b{}^\rho \right) + \frac{1}{\cal U} \bigl({\rm h}_a -  \frac{k_\lambda}{2}\,{\rm h}_a{}^\lambda \bigr) \, \bigl({\cal V}\,\tilde{{\cal G}}^{ab} -{\rm t}^a {\rm t}^b\bigr) \nonumber\\
& & \quad \quad  \times \bigl({\rm h}_b -  \frac{k_\rho}{2}\,{\rm h}_b{}^\rho \bigr) \, + \, \frac{1}{\, {\cal U}}\, \left({\cal U} \, \hat{\rm h}_\alpha{}^{0} + {\rmz}^\lambda \, \hat{\rm h}_{\alpha \lambda} \right) {\cal V}\,(\hat\kappa_{a\alpha\beta}\, {\rm t}^a)^{-1} \,\left({\cal U} \, \hat{\rm h}_\beta{}^{0} + {\rmz}^\rho \, \hat{\rm h}_{\beta \rho} \right)\biggr]\,, \\
& & \quad \quad + \frac{e^{3D}}{2\, \sqrt{\cal U}} \left[\left({\rm f}^0 \, {\rm h}_0 - {\rm f}^a\, {\rm h}_a \right) - \left({\rm f}^0\, {\rm h}^\lambda{}_0 - {\rm f}^a\, {\rm h}^\lambda{}_a \right)\, \frac{k_\lambda}{2} \right].\nonumber
\eea
where using the simplifications from the eqn. (\ref{eq:axionic-flux-nogo21}), the various non-zero ``axionic flux orbits" can be written out from the table \ref{tab_IIA-Fluxorbits} and those are simplified as under,
\bea
\label{eq:axionic-flux-nogo2}
& & \hskip-0.3cm {\rm f}_0  = e_0 + \, {\rm b}^a\, e_a + \frac{1}{2} \, \kappa_{abc} \, {\rm b}^a\, {\rm b}^b \,m^c + \frac{1}{6}\, \kappa_{abc}\,  {\rm b}^a \, {\rm b}^b\, {\rm b}^c \, m_0 \\
& & \hskip0.3cm - \, \xi^{0} \, ({\rm H}_0 + {\rm b}^a \, {w}_{a0}) - \, \xi^k \, ({\rm H}_k + {\rm b}^a \, {w}_{ak}) - {\xi}_\lambda \, ({\rm H}^\lambda + {\rm b}^a \, {w}_{a}{}^\lambda) \,, \nonumber\\
& & \hskip-0.3cm {\rm f}_a = e_a + \, \kappa_{abc} \,  {\rm b}^b \,m^c + \frac{1}{2}\, \kappa_{abc}\,  {\rm b}^b\, {\rm b}^c \, m_0 - \, \xi^{0} \, {w}_{a0} - \, \xi^k \, {w}_{ak} - {\xi}_\lambda \, {w}_a{}^\lambda\,, \nonumber\\
& & \hskip-0.3cm {\rm f}^a = m^a + m_0\,  {\rm b}^a \,, \quad  {\rm f}^0 = m_0\,, \nonumber\\
& & \nonumber\\
& & \hskip-0.3cm {\rm h}_0 = ({\rm H}_0 + {\rm b}^a \, {w}_{a0}) + {\rmz}^k \,({\rm H}_k + {\rm b}^a \, {w}_{ak}) + \, \frac{1}{2} \, \hat{k}_{\lambda mn} {\rmz}^m {\rmz}^n \, ({\rm H}^\lambda + {\rm b}^a \, {w}_{a}{}^\lambda), \nonumber\\
& & \hskip-0.3cm {\rm h}_{k0} = ({\rm H}_k + {\rm b}^a \, {w}_{ak}) +  \, \hat{k}_{\lambda k n}\, {\rmz}^n \, ({\rm H}^\lambda + {\rm b}^a \, {w}_{a}{}^\lambda), \quad {\rm h}^\lambda{}_0 = ({\rm H}^\lambda + {\rm b}^a \, {w}_{a}{}^\lambda)\,, \nonumber\\
& & \hskip-0.3cm {\rm h}_a = w_{a0} + {\rmz}^k \,w_{ak}  + \, \frac{1}{2} \, \hat{k}_{\lambda mn} {\rmz}^m {\rmz}^n \, w_a{}^\lambda, \quad {\rm h}_{ak} = w_{ak} +  \, \hat{k}_{\lambda k n}\, {\rmz}^n \,w_a{}^\lambda, \quad {\rm h}_a{}^\lambda = w_a{}^\lambda \,,\nonumber\\
& & \nonumber\\
& & \hskip-0.3cm \hat{\rm h}_{\alpha\lambda} = \hat{w}_{\alpha \lambda} + \hat{k}_{\lambda km} \, {\rmz}^m \, \hat{w}_\alpha{}^{k} - \frac{1}{2} \hat{k}_{\lambda km}  {\rmz}^k {\rmz}^m \hat{w}_\alpha{}^{0}, \quad \hat{h}_\alpha{}^{0} = \hat{w}_\alpha{}^{0}. \nonumber
\eea
Note that unlike the previous de-Sitter no-go scenario, now there can be non-trivial contributions generated from the $D$-terms via the geometric fluxes. Similar to the previous case, extracting the factor $\rho$ from the various volume moduli and metrics as in eqn. (\ref{eq:IIAmetric-rho}) the total scalar potential in eqn. (\ref{eq:main4IIA-nogo2gen}) simplifies to the following form,
\bea
\label{eq:main4IIA-nogo2}
& & \hskip-0.3cm V_{\rm IIA} = \frac{e^{4D}}{4\,\rho^3}\biggl[{\rm f}_0^2 + \rho^2\, {\rm f}_a \, \tilde{g}^{ab} \,{\rm f}_b + \rho^4\, {\rm f}^a \, \tilde{g}_{ab} \, {\rm f}^b + \rho^6\, ({\rm f}^0)^2\biggr] + \frac{e^{2D}}{4\,\rho^3}\biggl[\frac{{\rm h}_0^2}{\cal U} +\, \tilde{\cal G}^{ij}\,{\rm h}_{i0} \, {\rm h}_{j0} + \,\tilde{\cal G}_{\lambda \rho} {\rm h}^\lambda{}_0 \, {\rm h}^\rho{}_0 \biggr]\,\nonumber\\
& & \quad \qquad + \, \frac{e^{2D}}{4\,\rho\,}\biggl[\gamma^{a}\, \gamma^{b} \left(\frac{{\rm h}_a \, {\rm h}_b}{\cal U} + \, \tilde{\cal G}^{ij}\, {\rm h}_{ai} \, {\rm h}_{bj} \, + \,\tilde{\cal G}_{\lambda \rho}\, {\rm h}_a{}^\lambda\, {\rm h}_b{}^\rho \right) + \frac{1}{\cal U} \bigl({\rm h}_a -  \frac{k_\lambda}{2}\,{\rm h}_a{}^\lambda \bigr) \, \bigl(\tilde{g}^{ab} -\gamma^a \gamma^b\bigr) \nonumber\\
& & \quad \qquad  \times \bigl({\rm h}_b -  \frac{k_\rho}{2}\,{\rm h}_b{}^\rho \bigr) \, + \, \frac{1}{\, {\cal U}}\, \left({\cal U} \, \hat{\rm h}_\alpha{}^{0} + {\rmz}^\lambda \, \hat{\rm h}_{\alpha \lambda} \right) \,(\hat\kappa_{a\alpha\beta}\, \gamma^a)^{-1} \,\left({\cal U} \, \hat{\rm h}_\beta{}^{0} + {\rmz}^\rho \, \hat{\rm h}_{\beta \rho} \right)\biggr]\,, \\
& & \quad \qquad + \frac{e^{3D}}{2\, \sqrt{\cal U}} \left[\left({\rm f}^0 \, {\rm h}_0 - {\rm f}^a\, {\rm h}_a \right) - \left({\rm f}^0\, {\rm h}^\lambda{}_0 - {\rm f}^a\, {\rm h}^\lambda{}_a \right)\, \frac{k_\lambda}{2} \right].\nonumber
\eea
Now using the scalar potential in eqn. (\ref{eq:main4IIA-nogo2}) one can show that the following interesting relation holds,
\bea
& & \hskip-2cm \partial_D\,V_{\rm IIA} - \rho \, \partial_\rho V_{\rm IIA} = 3 \, V_{\rm IIA} + \frac{e^{2D}}{2\,\rho^3}\biggl[\frac{{\rm h}_0^2}{\cal U} +\, \tilde{\cal G}^{ij}\,{\rm h}_{i0} \, {\rm h}_{j0} + \,\tilde{\cal G}_{\lambda \rho} {\rm h}^\lambda{}_0 \, {\rm h}^\rho{}_0 \biggr] \nonumber\\
& & + \, \frac{e^{4D}}{4\,\rho^3}\biggl[4\, {\rm f}_0^2 + 2\,  \rho^2\, {\rm f}_a \, \tilde{g}^{ab} \,{\rm f}_b - 2\, \rho^6\, ({\rm f}^0)^2\biggr]\,.
\eea
One can observe the fact that for ${\rm f}^0 = m_0$ being set to zero, all the terms on the right hand side are non-negative which results in $(\partial_D\,V_{\rm IIA} - \rho \, \partial_\rho V_{\rm IIA}) \geq 3 \, V_{\rm IIA}$, and hence in this situation a new no-go condition holds despite of the fact that geometric fluxes are included. Moreover, one has the following inequality on the inflationary parameter $\epsilon$,
\bea
& & \epsilon \geq V_{\rm IIA}^{-2} \biggl[\frac{\rho^2}{3} {(\partial_\rho V_{\rm IIA})}^2 + \frac{1}{4} {(\partial_D V_{\rm IIA})}^2 \biggr] \\
& & \hskip0.3cm = V_{\rm IIA}^{-2} \biggl[\frac{1}{7} (3\, \partial_D\,V_{\rm IIA} - \rho \, \partial_\rho V_{\rm IIA})^2 + \frac{1}{84} (3\,\partial_D\,V_{\rm IIA} + 4\, \rho \, \partial_\rho V_{\rm IIA})^2 \biggr] \geq \frac{9}{7}\,. \nonumber
\eea
However it is also true that the earlier no-go condition is evaded with the simultaneous presence of geometric flux and the Romans mass term. The extremization conditions $\partial_D\,V_{\rm IIA} = 0 = \partial_\rho V_{\rm IIA}$ lead to the following form of the potential,
\bea
& & V_{\rm IIA}^{\rm ext}  = - \frac{e^{2D}}{6\,\rho^3}\biggl[\frac{{\rm h}_0^2}{\cal U} +\, \tilde{\cal G}^{ij}\,{\rm h}_{i0} \, {\rm h}_{j0} + \,\tilde{\cal G}_{\lambda \rho} {\rm h}^\lambda{}_0 \, {\rm h}^\rho{}_0 \biggr] - \frac{e^{4D}}{12\,\rho^3}\biggl[4\, {\rm f}_0^2 + 2\,  \rho^2\, {\rm f}_a \, \tilde{g}^{ab} \,{\rm f}_b - 2\, \rho^6\, ({\rm f}^0)^2\biggr]\,, \nonumber
\eea
which clearly opens up the possibility of getting de-Sitter via considering large enough value for the Romans mass parameter ${\rm f}^0 = m_0$ \cite{Haque:2008jz}. 

\subsection{$T$-dual de-Sitter no-go-2 in type IIB}
Now we want to know the $T$-dual version of this second type IIA no-go scenario on the type IIB side, and the $T$-duality from the non-zero fluxes in type IIA gives the flux ingredients of the type IIB setup as given in the table \ref{tab_no-go2}.
\noindent
\begin{table}[H]
\begin{center}
\begin{tabular}{|c||c|c|c|c||c|c|c|c|c|c||c|c|c|} 
\hline
& &&&&&& &&& &&&\\
IIA & $e_0 $ & $e_a$ & $m^a$ & $m_0$ & ${\rm H}_0$ & ${\rm H}_k$ & ${\rm H}^\lambda$ & ${w}_{a0}$ & ${w}_{ak}$ & ${w}_a{}^\lambda$ & $\hat{w}_{a}{}^0$ & $\hat{w}_a{}^k$ & $\hat{w}_{\alpha\lambda}$ \\
& &&&&&&&&&&&&\\
\hline
& &&&&&&&&&&&&\\
IIB & ${F}_0 $ & ${F}_i$ & ${F}^i$ & $- F^0$ & ${H}_0$ & ${\omega}_{a0}$ & $\hat{Q}^\alpha{}_0$ & ${H}_i$ & ${\omega}_{ai}$ & $\hat{Q}^\alpha{}_i$ & $-{R}_K$ & $- Q^a{}_K$ & $\hat{\omega}_{\alpha K}$ \\
& &&&&&&&&&&&&\\
\hline
\end{tabular}
\end{center}
\caption{Non-zero Type IIA fluxes and their respective $T$-duals for {\bf No-Go 2}.}
\label{tab_no-go2}
\end{table}
\noindent
It shows that for this scenario, the dual type IIB side can fairly get complicated with the presence of RR ($F_3$) flux along with all the (non-)geometric NS-NS fluxes unlike the type IIA case. Moreover, given the fact that this scenario corresponds to type IIA without any non-geometric flux, and therefore as we have analysed in previous section, this would be dual to type IIB with the `special solution' of Bianchi identities, in which half of the fluxes can be rotated away by a suitable symplectic transformation. Also, the Bianchi identities to worry about on type IIA and their dual type IIB side are simply the following ones,
\bea
& {\bf IIA:}& \quad {\rm H}^{\lambda} \, \hat{w}_{\alpha\lambda} = {\rm H}_{\hat{k}} \, \hat{w}_\alpha{}^{\hat{k}}, \qquad w_a{}^\lambda \, \hat{w}_{\alpha \lambda} = w_{a \hat{k}} \, \hat{w}_\alpha{}^{\hat{k}}\,;\\
& {\bf IIB:}& \quad  H_0 \, R_K + \omega_{a 0} \, Q^a{}_K + \hat{Q}^\alpha{}_0 \, \hat{\omega}_{\alpha K} = 0, \quad H_i \, R_K + \omega_{a i} \, Q^a{}_K + \hat{Q}^\alpha{}_i \, \hat{\omega}_{\alpha K} = 0\,. \nonumber
\eea
For implementing the `special solution' of Bianchi identities in the type IIB scalar potential, we need to switch-off the following axionic flux orbits,
\bea
& & h^0 = 0 = h^i, \quad h_a{}^i = 0 = h_a{}^0, \quad h^{\alpha i} = 0 = h^{\alpha 0}, \qquad \hat{h}_\alpha{}^K = 0 = \hat{h}^K\,,
\eea
where the last two hatted fluxes are parts of the $D$-term contributions. Using this simplification, and after a bit of reshuffling of terms, the dual scalar potential for the type IIB side can be subsequently read-off from the table \ref{tab_scalar-potential} and turns out to be given as under,
\bea
\label{eq:main4IIB-special}
& & V_{\rm IIB} = \frac{e^{4\phi}}{4\,{\cal V}^2\, {\cal U}}\biggl[f_0^2 + {\cal U}\, f^i \, {\cal G}_{ij} \, f^j + {\cal U}\, f_i \, {\cal G}^{ij} \,f_j + {\cal U}^2\, (f^0)^2\biggr]\,\\
& & \quad \qquad + \frac{e^{2\phi}}{4\,{\cal V}^2\,{\cal U}}\biggl[h_0^2  + \, {\cal V}\, {\cal G}^{ab}\,h_{a0} \, h_{b0} + \, {\cal V} \,{\cal G}_{\alpha \beta} \, h^\alpha{}_0 \, h^\beta{}_0 \nonumber\\
& & \quad \qquad + \, u^i\, u^j\, \left(h_i \, h_j + {\cal V} \,{\cal G}_{\alpha \beta}\,  h^\alpha{}_i\, h^\beta{}_j + {\cal V}\, {\cal G}^{ab}\, h_{ai} \, h_{bj} \right)\, \nonumber \\
& & \quad \qquad + \left({\cal U}\, {\cal G}^{ij} -  u^i\, u^j \right) \left(h_i - \frac{\ell_\alpha}{2} \,h^\alpha{}_i \right) \left(h_j - \frac{\ell_\beta}{2} \,h^\beta{}_j \right) \nonumber\\
& & \quad \qquad + \, {\cal U}\, \left({\cal V} \, \hat{h}_J{}^{0} - {t}^\alpha \, \hat{h}_{\alpha J} \right) \,(\hat\ell_{iJK}\, u^i)^{-1} \,\left({\cal V} \, \hat{h}_K{}^{0} - {t}^\beta \, \hat{h}_{\beta K} \right) \biggr] \nonumber\\
& & \quad \qquad + \frac{e^{3\phi}}{\, {\cal V}^2} \, \left[\left(f^0 \, h_0 - f^i\, h_i \right)\, - \left(f^0\, h^\alpha{}_0 - f^i\, h^\alpha{}_i \right)\, \frac{\ell_\alpha}{2} \right],\nonumber
\eea
where the simplified version of the non-trivial axionic flux orbits are given as below,
\bea
\label{eq:IIBorbits-nogo2}
& & \hskip-0.5cm f^0 = -\, F^0, \quad f^i = F^i - v^i\, F^0\,, \\
& & \hskip-0.5cm f_i = F_i +\, l_{ijk}\, v^j \, {F}^k\, - \frac{1}{2}\, l_{ijk}\, v^j\, v^k  \, {F}^0 - \omega_{ai} \, {c}^a - \hat{Q}^\alpha{}_i \, \hat{c}_\alpha   - \, c_0 \, h_i\,, \nonumber\\
& & \hskip-0.5cm f_0 = F_0 + v^i {F}_i + \frac{1}{2}\, l_{ijk} v^j v^k \, {F}^i - \frac{1}{6}\, l_{ijk} v^i v^j v^k \, {F}^0 - \omega_{a0} \, {c}^a - \hat{Q}^\alpha{}_0 \, \hat{c}_\alpha   - \, c_0 \, h_0\,, \nonumber\\
& & \nonumber\\
& & \hskip-0.5cm h_0 = H_0 + \omega_{a0} \, {b}^a + \frac{1}{2}\, \hat{\ell}_{\alpha a b}\, b^a b^b \, \hat{Q}^\alpha{}_0 + v^i \, h_i\, , \quad  h_i = H_i + \omega_{ai} \, {b}^a + \frac{1}{2}\, \hat{\ell}_{\alpha a b}\, b^a b^b \, \hat{Q}^\alpha{}_i\,,\nonumber\\
& &  \hskip-0.5cm h_{a0} = \omega_{a0} + \hat{Q}^\alpha{}_0 \hat{\ell}_{\alpha a b}\, b^b + v^i \, h_{ai}, \qquad h_{ai} = \omega_{ai} + \hat{Q}^\alpha{}_i \, \hat{\ell}_{\alpha a b}\, b^b. \nonumber \\
& & \hskip-0.5cm h^\alpha{}_0 = \hat{Q}^\alpha{}_0 + v^i \, \hat{Q}^\alpha{}_i\,\qquad h^\alpha{}_i = \hat{Q}^\alpha{}_i\,, \nonumber\\
& & \hskip-0.5cm \hat{h}_{\alpha K} = \hat{\omega}_{\alpha K}\, - Q^{a}{}_{K} \, \hat{\ell}_{\alpha a b} \, b^b +  \frac{1}{2}\hat{\ell}_{\alpha a b} \, b^a \,b^b\, R_K, \qquad \hat{h}_K{}^0 = -\, R_K\,. \nonumber
\eea
Now similar to the previous no-go-1 case, in order to prove that there is a new de-Sitter no-go scenario in type IIB side with non-geometric flux, all we need to do is to swap the role of complex-structure and the K\"ahler moduli. To see it explicitly we extract the $\sigma$ factor from the complex-structure moduli and the moduli space metrics as given in eqn. (\ref{eq:IIBmetric-sigma}). This leads to the type IIB scalar potential being written as under,
\bea
\label{eq:main4IIB-special}
& & V_{\rm IIB} = \frac{e^{4\phi}}{4\,{\cal V}^2\, \sigma^3}\biggl[f_0^2 + \sigma^2\, f_i \, {g}^{ij} \,f_j + \sigma^4\, f^i \, {g}_{ij} \, f^j + \sigma^6\, (f^0)^2 \biggr]\,\\
& & \quad \qquad + \frac{e^{2\phi}}{4\,{\cal V}^2\,\sigma^3}\biggl[\left(h_0^2  + \, {\cal V}\, {\cal G}^{ab}\,h_{a0} \, h_{b0} + \, {\cal V} \,{\cal G}_{\alpha \beta} \, h^\alpha{}_0 \, h^\beta{}_0 \right)\biggr] \nonumber\\
& & \quad \qquad + \frac{e^{2\phi}}{4\,{\cal V}^2\,\sigma} \biggl[\lambda^i\, \lambda^j\, \left(h_i \, h_j + {\cal V} \,{\cal G}_{\alpha \beta}\,  h^\alpha{}_i\, h^\beta{}_j + {\cal V}\, {\cal G}^{ab}\, h_{ai} \, h_{bj} \right)\, \nonumber \\
& & \quad \qquad + \,\left({g}^{ij} -  \lambda^i\, \lambda^j \right) \left(h_i - \frac{\ell_\alpha}{2} \,h^\alpha{}_i \right) \left(h_j - \frac{\ell_\beta}{2} \,h^\beta{}_j \right) \nonumber\\
& & \quad \qquad + \, \left({\cal V} \, \hat{h}_J{}^{0} - {t}^\alpha \, \hat{h}_{\alpha J} \right) \,(\hat\ell_{iJK}\, \lambda^i)^{-1} \,\left({\cal V} \, \hat{h}_K{}^{0} - {t}^\beta \, \hat{h}_{\beta K} \right) \biggr] \nonumber\\
& & \quad \qquad + \frac{e^{3\phi}}{2\, {\cal V}^2} \, \left[\left(f^0 \, h_0 - f^i\, h_i \right)\, - \left(f^0\, h^\alpha{}_0 - f^i\, h^\alpha{}_i \right)\, \frac{\ell_\alpha}{2} \right],\nonumber
\eea
where the angular moduli $\lambda^i$'s and the metrics $g^{ij}, g_{ij}$ do not have any dependence on the $\sigma$-modulus. Subsequently it is not hard to show that following relation holds,
\bea
& & \hskip-1cm \partial_\phi \,V_{\rm IIB} - \sigma \, \partial_\sigma V_{\rm IIB} = 3 V_{\rm IIB} + \frac{e^{2\phi}}{2\,{\cal V}^2\,\sigma^3}\biggl[\left(h_0^2  + \, {\cal V}\, {\cal G}^{ab}\,h_{a0} \, h_{b0} + \, {\cal V} \,{\cal G}_{\alpha \beta} \, h^\alpha{}_0 \, h^\beta{}_0 \right)\biggr] \nonumber\\
& & \hskip2cm+ \frac{e^{4\phi}}{4\,{\cal V}^2\, \sigma^3}\biggl[4\, f_0^2 + 2\, \sigma^2\, f_i \, {g}^{ij} \,f_j - 2\, \sigma^6\, (f^0)^2\biggr]\,.
\eea
The last term is the only non-positive term and this shows that for $f^0 \equiv - F^0 = 0$ we have the inequality $(\partial_\phi \,V_{\rm IIB} - \sigma \, \partial_\sigma V_{\rm IIB}) \geq 3 V_{\rm IIB}$. This immediately leads to a de-Sitter no-go theorem as at this extremum $\partial_\phi V_{\rm IIB} = 0 = \partial_\sigma V_{\rm IIB}$, the potential is allowed to take only the non-positive values as long as $f^0 = 0$ as we see below,
\bea
& & V_{\rm IIB}^{\rm ext}  = - \frac{e^{2\phi}}{3\,{\cal V}^2\,\sigma^3}\biggl[\left(h_0^2  + \, {\cal V}\, {\cal G}^{ab}\,h_{a0} \, h_{b0} + \, {\cal V} \,{\cal G}_{\alpha \beta} \, h^\alpha{}_0 \, h^\beta{}_0 \right)\biggr] \\
& & \hskip2cm - \frac{e^{4\phi}}{2\,{\cal V}^2\, \sigma^3}\biggl[4\, f_0^2 + 2\, \sigma^2\, f_i \, {g}^{ij} \,f_j - 2\, \sigma^6\, (f^0)^2\biggr]\,.\nonumber
\eea
Thus we are able to prove an interesting de-Sitter no-go theorem on type IIB side by $T$-dualizing the type IIA no-go, and moreover we have a possible way for finding de-Sitter by satisfying the necessary condition $F^0 \neq 0$ for the non-geometric flux with `special solutions'. 
\begin{mdframed}
\noindent
{\bf Type IIB No-Go theorem 2:} In type IIB framework with $O3/O7$ orientifold planes and (non-)geometric fluxes along with the standard $F_3, H_3$ fluxes, one cannot have stable de-Sitter minima with `special solutions' of Bianchi identities, unless $F^0$ component of the $F_3$ flux is non-zero, where $F_3 = F^\Lambda {\cal A}_\Lambda - F_\Lambda\, {\cal B}^\Lambda$, and $\Lambda \in \{0, 1,....,h^{2,1}_-\}$.
\end{mdframed}

\section{No-Go 3}
\label{sec_nogo3}
In the previous section, we have seen that after including the Romans mass term in type IIA or equivalently $F^0$ component of the three-form $F_3$-flux in type IIB, the necessary condition for getting the de-Sitter no-go is violated. This can be taken as a window to hunt for de-Sitter solutions. On the other hand, naively speaking, in order to restore the no-go condition or for finding another no-go, one would need to nullify the effects of these respective fluxes in the type IIA and the type IIB scenarios, and therefore one can ask the question if there are certain geometries which could be useful for this purpose. In this section we will show how the $K3$- or ${\mathbb T}^4$-fibred Calabi Yau threefolds could be useful in this regard as they facilitate a factorization in the moduli space as shown to be needed in \cite{Flauger:2008ad}.

\subsection{Type IIA with $K3$- or ${\mathbb T}^4$-fibred (CY) threefolds}
Superstring compactifications using $K3$- or ${\mathbb T}^4$-fibred CY threefolds present some interesting case as there is some kind of factorization guaranteed in the K\"ahler moduli space. By the theorem of \cite{Oguiso1993, Schulz:2004tt}, such a Calabi Yau threefold will have at least one two-cycle dual to a $K3$ or a ${\mathbb T}^4$ divisor which appears only linearly in the intersection polynomial\footnote{Such Calabi Yau threefolds with $K3/{\mathbb T}^4$-fibrations have been also useful for realizing Fibre inflation models \cite{Cicoli:2008gp, Cicoli:2016xae, Cicoli:2017axo}.}. In other words, the intersection numbers can be managed to split in the following manner by singling out a component through the splitting of index $a$ as $a = \{1, a'\}$,
\bea
\label{eq:intK3}
& & \hskip-1cm \kappa_{111} = 0 = \kappa_{11a'}, \quad \kappa_{1a'b'} \neq 0, \quad \hat{\kappa}_{1\alpha\beta} \neq 0, \quad  \hat{\kappa}_{a'\alpha\beta} = 0, \qquad {\rm where} \, \, a' \neq 1 \neq b'\,.
\eea
On top of that, in particular we also assume that $\kappa_{a'b'c'}=0$ and note that there is only one non-zero intersection of the type $\hat{\kappa}_{a\alpha\beta}$ with $a =1$. A concrete example of $K3$-fibred CY threefold with such even/odd splitting in the intersection numbers (and hence in the corresponding moduli space metrics) can be found in \cite{Gao:2013pra}. Recall that a non-zero intersection number of the type $\hat{\kappa}_{a\alpha\beta}$ is also essential for generating the $D$-terms by coupling through the (non-)geometric fluxes. 

Let us say that volume of a two-cycle which is singled out is denoted as ${\rm t}^1 = \rho_0$ leaving ${\rm t}^{a'}$ number of volume moduli as the remaining ones, and then the overall volume of the threefold can be written out as under,
\bea
& & {\cal V} = \frac{1}{6}\, \kappa_{abc}\, {\rm t}^a\, {\rm t}^b\, {\rm t}^c = \frac{1}{2}\, \kappa_{1a'b'}\, \rho_0\, {\rm t}^{a'}\, {\rm t}^{b'}\,,
\eea
which leaves the volume form as a homogeneous function of degree 2 in the the remaining prime-indexed K\"ahler moduli. Now we can still assume ${\rm t}^{a'} = \rho\, \gamma^{a'}$ where $\gamma^{a'}$'s are the remaining angular K\"ahler moduli satisfying $\kappa_{1a'b'}\gamma^{a'}\gamma^{b'} = 2$. This leads to a simple volume form given as under,
\bea
& & {\cal V} = \rho_0 \, \rho^2\,.
\eea
Before we come to the explicit detail on restoring the de-Sitter no-go condition by making an appropriate choice of the geometry, let us throw some more light on the motivation of looking at this $K3/{\mathbb T}4$-fibred geometry by considering the following Romans mass term as it appears in the type IIA scalar potential, 
\bea
& & \hskip0cm V_{{\rm f}^0} = \frac{e^{4D}}{2} {\cal V}\, ({\rm f}^0)^2 \,.
\eea
One can easily convince that using eqn. (\ref{eq:main4IIA-nogo2}) in which ${\cal V} = \rho^3$ simplification has been made we get the following relations,
\bea
& & (\partial_D V_{{\rm f}^0} - \rho\, \partial_\rho V_{{\rm f}^0}) = \, V_{{\rm f}^0}\, \Longrightarrow (\partial_D V_{\rm IIA} - \rho\, \partial_\rho V_{\rm IIA}) = 3 \,V_{\rm IIA} - 2\,V_{{\rm f}^0} + .....\,, 
\eea
where dots have some non-negative pieces as seen while deriving the no-go-2, and this way $V_{{\rm f}^0}$ appearing with a minus sign on the right hand side helps in evading the de-Sitter no-go condition. Now suppose we have a volume form of the type ${\cal V} = \rho_0\, \rho^2$ instead of ${\cal V} = \rho^3$, then the following relations hold,
\bea
\label{eq:factor-moduli-space-IIA}
& \hskip-1cm {\bf (I).} & \quad (\partial_D V_{{\rm f}^0} - \, \rho_0\, \partial_{\rho_0} V_{{\rm f}^0}) = 3\, V_{{\rm f}^0}\, \Longrightarrow (\partial_D V_{\rm IIA} - \rho_0\, \partial_{\rho_0} V_{\rm IIA}) = 3 \,V_{\rm IIA} + .....\,, \\
& \hskip-1cm {\bf (II).} & \quad (2\, \partial_D V_{{\rm f}^0} - \rho\, \partial_{\rho} V_{{\rm f}^0}) = 6\, V_{{\rm f}^0}\, \Longrightarrow (2\, \partial_D V_{\rm IIA} - \rho\, \partial_\rho V_{\rm IIA}) = 6 \,V_{\rm IIA} + .....\,, \nonumber
\eea
where we can see that now $V_{{\rm f}^0}$ can be completely absorbed in $V_{\rm IIA}$ and so negative piece with $V_{{\rm f}_0}$ is absent. Here we take an assumption (to be proven in a while) that one can appropriately make the flux choice to be such that all the other pieces inside the dots remain to be non-negative. Thus by considering these simple heuristics, one can anticipate to get another de-Sitter no-go with some appropriate choice of fluxes and geometries. 

Let us mention that one can also demand the splitting of intersection numbers on the mirror side, i.e. $k_{\lambda\rho\sigma}$ leading to the splitting in the complex structure moduli metric, to balance things from the $(\partial_D V_{{\rm f}^0})$ piece \cite{Flauger:2008ad} rather than considering $(\partial_\rho V_{{\rm f}^0})$ via taking a factorizable K\"ahler moduli space as we are considering. That may result in some new no-go scenarios, however we will not consider that case in this work.

To explore the details, using the choice for the triple-intersection numbers given in eqn. (\ref{eq:intK3}) and the definitions of the metric given in table \ref{tab_scalar-potential} we have the following block-diagonal forms for the (inverse-)moduli space metrics,
\bea
\label{eq:K3-moduli-space-IIA}
& & \hskip-1cm {\cal V} \,\tilde{\cal G}^{ab} = \begin{pmatrix} 
\rho_0^2 & \quad 0\\
0 & \,\, \, \,\rho^2 \left(\gamma^{a'}\, \gamma^{b'} - \tilde\kappa^{a'b'}\right)
\end{pmatrix}, \quad {\cal V}\,\tilde{\cal G}_{ab} = \begin{pmatrix} 
\rho^{4} & \quad 0\\
0 & \quad \rho_0^{2}\,\rho^{2} \left(\tilde\kappa_{a'}\tilde\kappa_{b'} - \tilde\kappa_{a'b'}\right) 
\end{pmatrix}\,,
\eea
where $a' \in \{2, 3, .., h^{1,1}_-\}$ and the angular quantities with $a'$ indices do not depend on any of the moduli $\rho_0$ and $\rho$. From the scalar potential in eqn. (\ref{eq:main4IIA-nogo2gen}), which is relevant for this type IIA case with geometric flux, we observe that the volume moduli $\rho_0$ and $\rho$ can appear through factors like $({\cal V} \, \tilde{\cal G}^{ab}), ({\cal V} \, \tilde{\cal G}_{ab}), ({\rm t}^a\, {\rm t}^b)$ or $(\hat\kappa_{a\alpha\beta}{\rm t}^a)$. As we have seen from eqn. (\ref{eq:K3-moduli-space-IIA}), the moduli space metrics are already block diagonal with the splitting of index `$a$' as $a = \{1, a'\}$. Also note that the piece with $(\hat\kappa_{a\alpha\beta}{\rm t}^a)^{-1}$ will only depend on $\rho_0$ (and not on the $\rho$) modulus as we have assumed in eqn. (\ref{eq:intK3}) that $\hat\kappa_{1\alpha\beta}$ is the only non-zero intersection with index $\alpha, \beta$ being in the even $(1,1)$-cohomology. However, scalar potential pieces involving the factor $({\rm t}^a\, {\rm t}^b)$ can generate off-diagonal mixings and so might disturb the balance of pieces in $(\partial_D V_{\rm IIA} - \rho_0\, \partial_{\rho_0} V_{\rm IIA}) = 3\,V_{\rm IIA} + ...$, so that to keep retaining the pieces hidden in the dots as positive semi-definite, something which was established for the earlier no-go-2. To concretize these arguments, we simplify the geometric type IIA scalar potential given in eqn. (\ref{eq:main4IIA-nogo2gen}) utilising the above splitting of the moduli space metrics, and it turns out to be given as under,
\bea
\label{eq:main4IIA-nogo3}
& & \hskip-0.3cm V_{\rm IIA} = \frac{e^{4D}}{4\,\rho_0\, \rho^2}\biggl[({\rm f}_0)^2 + \rho_0^2\, ({\rm f}_1)^2 + \rho^2\, {\rm f}_{a'} \, (\gamma^{a'}\, \gamma^{b'} - \tilde\kappa^{a'b'}) \,{\rm f}_{b'}  \\
& & \quad \quad + \, \rho^4 ({\rm f}^1)^2 + \, \rho_0^{2}\,\rho^{2} \, {\rm f}^{a'} \, (\tilde\kappa_{a'}\tilde\kappa_{b'} - \tilde\kappa_{a'b'}) \, {\rm f}^{b'} + \rho_0^2\, \rho^4\, ({\rm f}^0)^2\biggr] \nonumber\\
& & \quad \quad + \frac{e^{2D}}{4\,\rho_0\, \rho^2}\biggl[\frac{{\rm h}_0^2}{\cal U} +\, \tilde{\cal G}^{ij}\,{\rm h}_{i0} \, {\rm h}_{j0} + \,\tilde{\cal G}_{\lambda \rho} {\rm h}^\lambda{}_0 \, {\rm h}^\rho{}_0 \biggr]\,\nonumber\\
& & \quad \quad + \, \frac{e^{2D}}{4\, \rho^2} \times \rho_0 \biggl[\left(\frac{{\rm h}_1 \, {\rm h}_1}{\cal U} + \, \tilde{\cal G}^{ij}\, {\rm h}_{1i} \, {\rm h}_{1j} \, + \,\tilde{\cal G}_{\lambda \rho}\, {\rm h}_1{}^\lambda\, {\rm h}_1{}^\rho \right) \biggr] \nonumber\\
& & \quad \quad + \, \frac{e^{2D}}{2\, \rho\,}\biggl[\gamma^{a'} \left(\frac{{\rm h}_{a'} \, {\rm h}_1}{\cal U} + \, \tilde{\cal G}^{ij}\, {\rm h}_{a'i} \, {\rm h}_{1j} \, + \,\tilde{\cal G}_{\lambda \rho}\, {\rm h}_{a'}{}^\lambda\, {\rm h}_1{}^\rho \right) \biggr] \nonumber\\
& & \quad \quad + \, \frac{e^{2D}}{4\,\rho_0}\biggl[\gamma^{a'}\, \gamma^{b'} \left(\frac{{\rm h}_{a'} \, {\rm h}_{b'}}{\cal U} + \, \tilde{\cal G}^{ij}\, {\rm h}_{a'i} \, {\rm h}_{b'j} \, + \,\tilde{\cal G}_{\lambda \rho}\, {\rm h}_{a'}{}^\lambda\, {\rm h}_{b'}{}^\rho \right) \biggr] \nonumber\\
& & \quad \quad - \, \frac{e^{2D}}{4\,\rho_0\,{\cal U}}\biggl[\bigl({\rm h}_{a'} -  \frac{k_\lambda}{2}\,{\rm h}_{a'}{}^\lambda \bigr) \, \tilde\kappa^{a'b'} \bigl({\rm h}_{b'} -  \frac{k_\rho}{2}\,{\rm h}_{b'}{}^\rho \bigr) \biggr]\, \nonumber\\
& & \quad \quad - \, \frac{e^{2D}}{2\, \rho\,{\cal U}}\biggl[\bigl({\rm h}_1 -  \frac{k_\lambda}{2}\,{\rm h}_1{}^\lambda \bigr)\, \gamma^{a'} \bigl({\rm h}_{a'} -  \frac{k_\rho}{2}\,{\rm h}_{a'}{}^\rho \bigr) \biggr]\, \nonumber\\
& & \quad \quad + \, \frac{e^{2D}}{4\,\rho_0\, {\cal U}}\biggl[\left({\cal U} \, \hat{\rm h}_\alpha{}^{0} + {\rmz}^\lambda \, \hat{\rm h}_{\alpha \lambda} \right) \,(\hat\kappa_{1\alpha\beta}\, \gamma^1)^{-1} \,\left({\cal U} \, \hat{\rm h}_\beta{}^{0} + {\rmz}^\rho \, \hat{\rm h}_{\beta \rho} \right)\biggr]\,\nonumber\\
& & \quad \quad + \frac{e^{3D}}{2\, \sqrt{\cal U}} \left[\left({\rm f}^0 \, {\rm h}_0 - {\rm f}^a\, {\rm h}_a \right) - \left({\rm f}^0\, {\rm h}^\lambda{}_0 - {\rm f}^a\, {\rm h}^\lambda{}_a \right)\, \frac{k_\lambda}{2} \right],\nonumber
\eea
where the flux orbits can be read-off from the eqn. (\ref{eq:axionic-flux-nogo2}) after imposing the splitting of indices as $a = \{1, a'\}$ and using the intersection numbers given in eqn. (\ref{eq:intK3}). Now from this complicated potential we can see the off-diagonal mixing e.g. arising from $({\rm t}^a\, {\rm t}^b)$ factor as we discussed before. This issue can be avoided by appropriately setting the respective fluxes coupled in the off-diagonal blocks to zero. i.e. by taking either of the following two cases which subsequently leads to the new de-Sitter no-go scenarios,
\bea
\label{eq:IIA-nogo3}
& \hskip-0.5cm {\bf (I).} & \quad {\rm h}_1 = {\rm h}_{1k} = {\rm h}_1{}^\lambda = 0 \quad \Longleftrightarrow \quad {w}_{10} = {w}_{1k} = {w}_1{}^\lambda = 0\,, \\
& & \hskip2cm  \Longrightarrow \quad (\partial_D V_{\rm IIA} - \rho_0\, \partial_{\rho_0} V_{\rm IIA}) \geq 3 \,V_{\rm IIA};\nonumber\\
& \hskip-0.5cm {\bf (II).} & \quad {\rm h}_{a'0} = {\rm h}_{a'k} = {\rm h}_{a'}{}^\lambda = \hat{\rm h}_{\alpha}{}^0 = \hat{\rm h}_{\alpha}{}^k = \hat{\rm h}_{\alpha \lambda}= 0 \Longleftrightarrow \nonumber\\
& & \hskip0cm  {w}_{a'0} = {w}_{a'k} = {w}_{a'}{}^\lambda = \hat{w}_{\alpha}{}^0 = \hat{w}_{\alpha}{}^k = \hat{w}_{\alpha \lambda}= 0\,  \Longrightarrow  (2 \partial_D V_{\rm IIA} - \rho\, \partial_{\rho} V_{\rm IIA}) \geq 6 \,V_{\rm IIA} . \nonumber
\eea
Also note that in the no-go scenarios corresponding to the above two cases, one has to impose those extra flux conditions about vanishing of certain fluxes to determine the simplified axionic flux orbits from their generic expressions. However given their nature of being independent of the saxion, it doesn't bother us for our purpose as we are only interested in considering the saxionic derivatives of the potential to look for the possible no-go inequalities. 

\subsection{$T$-dual de-Sitter no-go-3 in type IIB}
On the lines of computations done for the explicit $T$-dualization of the two de-Sitter no-go scenarios, one can be convinced that the no-go-3 in (\ref{eq:IIA-nogo3}) can be easily $T$-dualized to find new no-go scenarios on the type IIB side. For this to happen, the assumption to make is that type IIB compactification should be done on the CY threefolds which have $K3$- or ${\mathbb T}^4$-fibred mirror CYs. So this framework should not be confused with having type IIB compactification on the $K3$- or ${\mathbb T}^4$-fibred CY itself, although there might be different set of no-go's for that case, but those would not be the ones we are considering as type IIA no-go-3.  

Having said the above, now the complex structure side can be studied by the mirror CY, and hence will inherit the splitting of complex-structure moduli space on the type IIB side such that one can single out two complex structure moduli $\sigma_0$ and $\sigma$ such that,
\bea
\label{eq:K3-moduli-space-IIB}
& & u^1 = \sigma_0, \qquad u^{i'} = \sigma\, \lambda^{i'}, \qquad  l_{1i'j'}\, \lambda^{i'}\, \lambda^{j'} = 2\,, \qquad {\cal U} = \sigma_0 \, \sigma^2\,,\\
& & \hskip-1cm {\cal U} \,{\cal G}^{ij} = \begin{pmatrix} 
\sigma_0^2 & \quad 0\\
0 & \,\, \, \,\sigma^2 (\lambda^{i'}\, \lambda^{j'} - \tilde{l}^{i'j'})
\end{pmatrix}, \quad {\cal U}\,{\cal G}_{ij} = \begin{pmatrix} 
\sigma^{4} & \quad 0\\
0 & \quad \sigma_0^{2}\,\sigma^{2} (\tilde{l}_{i'}\tilde{l}_{j'} - \tilde{l}_{i'j'}) 
\end{pmatrix}\,, \nonumber
\eea
where the indices $i'$'s denote the remaining complex structure moduli different from $u^1$ and quantities like $\tilde{l}_i$ etc. are the ones which only depend on the angular complex structure moduli. Under these circumstances, the type IIB scalar potential can be explicitly given as under,
\bea
\label{eq:IIBpotential-nogo3}
& & V_{\rm IIB} = \frac{e^{4\phi}}{4\,{\cal V}^2\, \sigma_0 \, \sigma^2}\biggl[(f_0)^2 + \left(\sigma^4\, (f^1)^2 + \sigma_0^2\, \sigma^2 \, f^{i'} \, (\tilde{l}_{i'}\tilde{l}_{j'} - \tilde{l}_{i'j'})\, f^{j'} \right) \\
& & \quad \qquad + \left(\sigma_0^2 \, (f_1)^2 + \sigma^2\, f_{i'} \, (\lambda^{i'}\, \lambda^{j'} - \tilde{l}^{i'j'}) \,f_{j'} \right)+ \sigma_0^2 \, \sigma^4\, (f^0)^2\biggr]\,\nonumber\\
& & \quad \qquad + \frac{e^{2\phi}}{4\,{\cal V}^2\,\sigma_0 \, \sigma^2}\biggl[h_0^2  + \, {\cal V}\, {\cal G}^{ab}\,h_{a0} \, h_{b0} + \, {\cal V} \,{\cal G}_{\alpha \beta} \, h^\alpha{}_0 \, h^\beta{}_0 \nonumber\\
& & \quad \qquad + \, \sigma_0^2\, \left((h_1)^2 + {\cal V} \,{\cal G}_{\alpha \beta}\,  h^\alpha{}_1\, h^\beta{}_1 + {\cal V}\, {\cal G}^{ab}\, h_{a1} \, h_{b1} \right)\, \nonumber \\
& & \quad \qquad +\, \sigma^2\, \lambda^{i'}\, \lambda^{j'}\, \left(h_{i'} \, h_{j'} + {\cal V} \,{\cal G}_{\alpha \beta}\,  h^\alpha{}_{i'}\, h^\beta{}_{j'} + {\cal V}\, {\cal G}^{ab}\, h_{ai'} \, h_{bj'} \right)\, \nonumber \\
& & \quad \qquad + \sigma^2 (\lambda^{i'}\, \lambda^{j'} -  \tilde{l}^{i'j'}) \, \left(h_{i'} - \frac{\ell_\alpha}{2} \,h^\alpha{}_{i'} \right) \left(h_{j'} - \frac{\ell_\beta}{2} \,h^\beta{}_{j'} \right) \nonumber\\
& & \quad \qquad + \, \sigma^2\, \left({\cal V} \, \hat{h}_J{}^{0} - {t}^\alpha \, \hat{h}_{\alpha J} \right) \,(\hat\ell_{1JK})^{-1} \,\left({\cal V} \, \hat{h}_K{}^{0} - {t}^\beta \, \hat{h}_{\beta K} \right) \biggr] \nonumber\\
& & \quad \qquad + \frac{e^{3\phi}}{\, {\cal V}^2} \, \left[\left(f^0 \, h_0 - f^i\, h_i \right)\, - \left(f^0\, h^\alpha{}_0 - f^i\, h^\alpha{}_i \right)\, \frac{\ell_\alpha}{2} \right],\nonumber
\eea
where the only interest for us at the moment lies in the saxionic moduli $\sigma_0$ and $\sigma$, though for completion we do provide the explicit expressions for all the axionic flux orbits as under,
\bea
\label{eq:IIBorbits-nogo3}
& & \hskip-0.5cm f^0 = -\, F^0, \quad f^1 = F^1 - v^1\, F^0\,, \quad f^{i'} = F^{i'} - v^{i'}\, F^0,\\
& & \hskip-0.5cm f_1 = F_1 +\, l_{1i'j'}\, v^{i'} \, {F}^{j'}\, - \frac{1}{2}\, l_{1j'k'}\, v^{j'}\, v^{k'}  \, {F}^0 - \omega_{a1} \, {c}^a - \hat{Q}^\alpha{}_1 \, \hat{c}_\alpha   - \, c_0 \, h_1\,, \nonumber\\
& & \hskip-0.5cm f_{i'} = F_{i'} +\, l_{1i'j'}\, (v^{j'} \, {F}^{1} + v^{1} \, {F}^{j'})\, - \, l_{1i'j}\, v^{1}\, v^{j'}  \, {F}^0 - \omega_{a{i'}} \, {c}^a - \hat{Q}^\alpha{}_{i'} \, \hat{c}_\alpha   - \, c_0 \, h_{i'}\,, \nonumber\\
& & \hskip-0.5cm f_0 = F_0  + v^1 {F}_1+ v^{i'} {F}_{i'} + \frac{1}{2}\, l_{1i'j'} v^{i'} v^{j'} \, {F}^1 + l_{1i'j'} v^1 v^{i'} \, {F}^{j'} \nonumber\\
& & \hskip1cm - \frac{1}{2}\, l_{1i'j'} v^{i'} v^{j'} v^1 \, {F}^0 - \omega_{a0} \, {c}^a - \hat{Q}^\alpha{}_0 \, \hat{c}_\alpha   - \, c_0 \, h_0\,, \nonumber\\
& & \nonumber\\
& & \hskip-0.5cm h_0 = H_0 + \omega_{a0} \, {b}^a + \frac{1}{2}\, \hat{\ell}_{\alpha a b}\, b^a b^b \, \hat{Q}^\alpha{}_0 + v^1 \, h_1\,+ v^{i'} \, h_{i'} , \nonumber\\
& & \hskip-0.5cm h_1 = H_1 + \omega_{a1} \, {b}^a + \frac{1}{2}\, \hat{\ell}_{\alpha a b}\, b^a b^b \, \hat{Q}^\alpha{}_1\,, \qquad h_{i'} = H_{i'} + \omega_{ai'} \, {b}^a + \frac{1}{2}\, \hat{\ell}_{\alpha a b}\, b^a b^b \, \hat{Q}^\alpha{}_{i'}\,,\nonumber\\
& &  \hskip-0.5cm h_{a0} = \omega_{a0} + \hat{Q}^\alpha{}_0 \hat{\ell}_{\alpha a b}\, b^b + v^1 \, h_{a1} + v^{i'} \, h_{ai'}, \qquad h_{a1} = \omega_{a1} + \hat{Q}^\alpha{}_1 \, \hat{\ell}_{\alpha a b}\, b^b,  \nonumber \\
& & \hskip-0.5cm h_{ai'} = \omega_{ai'} + \hat{Q}^\alpha{}_{i'} \, \hat{\ell}_{\alpha a b}\, b^b, \quad h^\alpha{}_0 = \hat{Q}^\alpha{}_0 + v^1 \, \hat{Q}^\alpha{}_1 + v^{i'} \, \hat{Q}^\alpha{}_{i'}\,, \, \, \, h^\alpha{}_1 = \hat{Q}^\alpha{}_1\,, \, \, \, h^\alpha{}_{i'} = \hat{Q}^\alpha{}_{i'}, \nonumber\\
& & \hskip-0.5cm \hat{h}_{\alpha K} = \hat{\omega}_{\alpha K}\, - Q^{a}{}_{K} \, \hat{\ell}_{\alpha a b} \, b^b +  \frac{1}{2}\hat{\ell}_{\alpha a b} \, b^a \,b^b\, R_K, \qquad \hat{h}_K{}^0 = -\, R_K\,. \nonumber
\eea
A close look at the scalar potential in eqn. (\ref{eq:IIBpotential-nogo3}) confirms that one can have the following two $T$-dual cases,
\bea
& \hskip-1cm {\bf (I).} & \quad {h}_1 = {h}_{a1} = {h}^\alpha{}_1 = 0 \quad \Longleftrightarrow \quad {H}_{1} = {\omega}_{a1} = \hat{Q}^\alpha{}_1 = 0\,,\\
& & \hskip2cm \Longrightarrow \quad (\partial_D V_{\rm IIB} - \sigma_0\, \partial_{\sigma_0} V_{\rm IIB}) \geq 3 \,V_{\rm IIA}\,,\nonumber\\
& \hskip-1cm {\bf (II).} & \quad {h}_{i'} = {h}_{ai'} = {h}^\alpha{}_{i'} = \hat{h}_{\alpha K} = h^a{}_K = \hat{h}_{K}{}^0 = 0 \nonumber\\
& & \quad \Longleftrightarrow \quad H_{i'}= \omega_{ai'} = \hat{Q}^\alpha{}_{i'} = \hat{\omega}_{\alpha K} = Q^a{}_K = R_K = 0\, \quad \nonumber\\
& & \hskip2cm \Longrightarrow \quad (2\, \partial_D V_{\rm IIB} - \sigma\, \partial_\sigma V_{\rm IIB}) \geq 6 \,V_{\rm IIB}, \nonumber
\eea
This result can be summarized in the following no-go condition.
\begin{mdframed}
\noindent
{\bf Type IIB No-Go theorem 3:} In type IIB framework with $O3/O7$ orientifold planes and (non-)geometric fluxes along with the standard $F_3, H_3$ fluxes, one cannot have stable de-Sitter minima with `special solutions' of Bianchi identities, if the complex structure moduli space exhibit a factorization on top of suitably having some of the flux components set to zero. This can happen when the mirror of the type IIB compactifying CY is a particular type of $K3/{\mathbb T}^4$-fibred CY threefold satisfying eqn. (\ref{eq:intK3}).
\end{mdframed} 
\noindent

\subsection{More de-Sitter no-go conditions for toroidal examples}
This no-go 3 appears to be rather a complicated statement to make, however it has several interesting implications. To illustrate what it means in a simple way, we consider the toroidal models based on type IIA and type IIB compactifications using orientifold of the ${\mathbb T}^6/({\mathbb Z}_2 \times {\mathbb Z}_2)$ orbifold. To being with, let us mention that this no-go 3 can be applied directly to these conventional vanilla toroidal orientifold models which have been studied numerous number of times. This model has the only intersection number non-zero to be,
\bea
& \hskip-1cm {\rm IIA:} & \qquad \kappa_{123} = 1, \quad \ \qquad  k_{123} = 1\,, \\
& \hskip-1cm {\rm IIB:} & \qquad \, \ell_{123} = 1, \quad \qquad \, \, \, \, l_{123} = 1\,, \nonumber
\eea
while all the other intersection numbers are  zero. With the standard orientifold involution there are no $D$-terms present in type IIA or type IIB settings. So the total scalar potential arise from the $F$-term contributions itself. In addition, let us note that the even $(1,1)$-cohomology is trivial in type IIA while the odd $(1,1)$-cohomology is trivial in type IIB implying that fluxes/moduli with indices $k$ in type IIA and with index $a$ in type IIB are absent. 

\subsubsection*{type IIA}
It turns out that 12 axionic flux orbits are identically zero in this construction, which in addition does not include non-geometric ${\rm Q}$ and ${\rm R}$ fluxes,
\bea
& & \hskip-1cm {\rm h}^a = 0 = {\rm h}^0, \quad  {\rm h}_{k0} = {\rm h}_{ak}  = {\rm h}^a{}_k = {\rm h}_k{}^0 = 0, \quad  {\rm h}^{a\lambda} = 0 =  {\rm h}^{\lambda 0}\,, \\
& & \qquad \quad \hat{{\rm h}}_{\alpha}{}^{0} = \hat{{\rm h}}_{\alpha\lambda}  = \hat{{\rm h}}^{\alpha0} = \hat{{\rm h}}^\alpha{}_\lambda = 0\,.\nonumber
\eea
As there can be equivalence between the three ${\mathbb T}^2$'s appearing in the six-torus, and therefore one can single out $\rho_0$ modulus from any of the three $t^a$'s, say we take $t^1 = \rho_0$ and subsequently the remaining $2 \times 2$ sector in the K\"ahler moduli space is block diagonal. In fact it is completely diagonal in all the three moduli, though we need it only partially. Noting that the only fluxes which can get non-zero values in this model are the followings:
\bea
& & {\rm h}_0, \qquad {\rm h}_a, \qquad {\rm h}_{a}{}^\lambda, \qquad {\rm h}^\lambda{}_0,
\eea
our no-go-3 implies that one would end up in having de-Sitter no-go scenarios if one switches-off certain fluxes as mentioned in table \ref{tab_IIA-nogo3}.
\noindent
\begin{table}[H]
\begin{center}
\begin{tabular}{|c|c|c|} 
\hline
& &\\
${\rm h}_1 = {\rm h}_1{}^\lambda = 0$ & ${w}_{10} = {w}_1{}^\lambda = 0$ & $\partial_D V_{\rm IIA} - \rho_0 \partial_{\rho_0} V_{\rm IIA} \geq 3V_{\rm IIA}$ \\
${\rm h}_2 = {\rm h}_2{}^\lambda = 0$ & ${w}_{20} = {w}_2{}^\lambda = 0$ & $\partial_D V_{\rm IIA} - \rho_0 \partial_{\rho_0} V_{\rm IIA} \geq 3 V_{\rm IIA}$ \\
${\rm h}_3 = {\rm h}_3{}^\lambda = 0$ & ${w}_{30} = {w}_3{}^\lambda = 0$ & $\partial_D V_{\rm IIA} - \rho_0 \partial_{\rho_0} V_{\rm IIA} \geq 3V_{\rm IIA}$ \\
& &\\
\hline
& & \\
${\rm h}_{20} = {\rm h}_{30} = {\rm h}_{2}{}^\lambda = {\rm h}_{3}{}^\lambda = 0$ & ${w}_{20} = {w}_{30} = {w}_{2}{}^\lambda = {w}_{3}{}^\lambda = 0$ & $2 \partial_D V_{\rm IIA} - \rho \partial_{\rho} V_{\rm IIA} \geq 6 V_{\rm IIA}$\\
${\rm h}_{10} = {\rm h}_{30} = {\rm h}_{1}{}^\lambda = {\rm h}_{3}{}^\lambda =0 $ & ${w}_{10} = {w}_{30} = {w}_{1}{}^\lambda = {w}_{3}{}^\lambda = 0$ & $2 \partial_D V_{\rm IIA} - \rho \partial_{\rho} V_{\rm IIA} \geq 6 V_{\rm IIA}$\\
${\rm h}_{10} = {\rm h}_{20} = {\rm h}_{1}{}^\lambda = {\rm h}_{2}{}^\lambda = 0$ & ${w}_{10} = {w}_{20} = {w}_{1}{}^\lambda = {w}_{2}{}^\lambda = 0$ & $2 \partial_D V_{\rm IIA} - \rho \partial_{\rho} V_{\rm IIA} \geq 6 V_{\rm IIA}$\\
\hline
\end{tabular}
\end{center}
\caption{Type IIA de-Sitter no-go scenarios with ${\mathbb T}^6/({\mathbb Z}_2 \times {\mathbb Z}_2)$ having geometric flux.}
\label{tab_IIA-nogo3}
\end{table}
\noindent
The particular models of table \ref{tab_IIA-nogo3} present those cases in which one would have de-Sitter no-go irrespective of the fact whether  the Romans mass term is zero or non-zero. This simply means that these are the examples in which geometric fluxes are not enough to evade the no-go-2 despite having non-zero Romans mass. Moreover, from the observations from table \ref{tab_IIA-nogo3} it is not hard to guess that if all the geometric fluxes are zero, one gets back to the no-go-1 having an inequality of the type: $(3 \partial_D V_{\rm IIA} - \rho \partial_{\rho} V_{\rm IIA}) \geq 9 \,V_{\rm IIA}$ !

\subsubsection*{type IIB}
Now an interesting question to ask is what happens to the dual type IIB side which would involve non-geometric fluxes as well, unlike the type IIA case. It turns out that 12 axionic flux orbits are identically zero in this construction, and they are given as under:
\bea
& & h^0 = h^i = 0, \qquad h_{a0} = h_{ai} = h_a{}^i = h_a{}^0 = 0, \qquad h^{\alpha i} = h^{\alpha 0} = 0, \\
& & \hat{h}_K = \hat{h}_{\alpha K} = \hat{h}_\alpha{}^K = \hat{h}^K = 0\,.\nonumber
\eea
Now due to symmetries in the intersection number $l_{ijk}$, one can single out a $\sigma_0$ modulus from any of the three complex structure saxions $u^i$'s, say we take $u^1 = \sigma_0$ and subsequently the remaining $2 \times 2$ sector in the complex structure moduli space is block diagonal, and one can write ${\cal U} = \sigma_0\, \sigma^2$. As before, it is completely diagonal in all the three moduli. Noting that the only fluxes which can get non-zero values in this model are the following ones:
\bea
& & {h}_0, \qquad {h}_i, \qquad {h}^{\alpha}{}^0, \qquad {h}^\alpha{}_i,
\eea
our no-go-3 implies that one would end up in having de-Sitter no-go scenarios if one switches-off certain fluxes as mentioned in table \ref{tab_IIB-nogo3}.
\noindent
\begin{table}[H]
\begin{center}
\begin{tabular}{|c|c|c|} 
\hline
& &\\
${h}_1 = {h}^\alpha{}_1 = 0$ & ${H}_{1} = \hat{Q}^\alpha{}_1 = 0$ & $(\partial_\phi V_{\rm IIB} - \sigma_0 \partial_{\sigma_0} V_{\rm IIB}) \geq 3 \,V_{\rm IIB}$ \\
${h}_2 = {h}^\alpha{}_2 = 0$ & ${H}_{2} = \hat{Q}^\alpha{}_2 = 0$ & $(\partial_\phi V_{\rm IIB} - \sigma_0 \partial_{\sigma_0} V_{\rm IIB}) \geq 3 \,V_{\rm IIB}$ \\
${h}_3 = {h}^\alpha{}_3 = 0$ & ${H}_{3} = \hat{Q}^\alpha{}_3 = 0$ & $(\partial_\phi V_{\rm IIB} - \sigma_0 \partial_{\sigma_0} V_{\rm IIB}) \geq 3 \,V_{\rm IIB}$ \\
& &\\
\hline
& & \\
${h}_2 = {h}_3 = {h}^\alpha{}_2 = {h}^\alpha{}_3 = 0$ & ${H}_{2} = {H}_{3} = \hat{Q}^\alpha{}_2 =  \hat{Q}^\alpha{}_3 =0$ & $(2 \partial_\phi V_{\rm IIB} - \sigma \partial_{\sigma} V_{\rm IIB}) \geq 6 \,V_{\rm IIB}$\\
${h}_3 = {h}_1 = {h}^\alpha{}_3 = {h}^\alpha{}_1 = 0$ & ${H}_{3} = {H}_{1} = \hat{Q}^\alpha{}_3 =  \hat{Q}^\alpha{}_1 =0$ & $(2 \partial_\phi V_{\rm IIB} - \sigma \partial_{\sigma} V_{\rm IIB}) \geq 6 \,V_{\rm IIB}$\\
${h}_1 = {h}_2 = {h}^\alpha{}_1 = {h}^\alpha{}_2 = 0$ & ${H}_{1} = {H}_{2} = \hat{Q}^\alpha{}_1 =  \hat{Q}^\alpha{}_2 =0$ & $(2 \partial_\phi V_{\rm IIB} - \sigma \partial_{\sigma} V_{\rm IIB}) \geq 6 \,V_{\rm IIB}$\\
\hline
\end{tabular}
\end{center}
\caption{Type IIB de-Sitter no-go scenarios with ${\mathbb T}^6/({\mathbb Z}_2 \times {\mathbb Z}_2)$ having (non-)geometric fluxes.}
\label{tab_IIB-nogo3}
\end{table}
\noindent
The particular models of table \ref{tab_IIB-nogo3} present those cases in which one would have de-Sitter no-go irrespective of the fact whether the $F^0$ components of the RR $F_3$ flux is zero or non-zero, and moreover despite having some non-geometric fluxes being turned-on. This means that these are the examples in which non-geometric fluxes are not enough to evade the no-go-2 due to the presence of some specific geometries inherited from the six-torus. 

\section{Summary and conclusions}
\label{sec_conclusions}
In this article, we have $T$-dualized several de-Sitter no-go scenarios which have been well known in the type IIA flux compactifications since more than a decade. This subsequently leads to a set of peculiar de-Sitter no-go scenarios in the type IIB flux compactifications with (non-)geometric fluxes. 

Before exploring the de-Sitter no-go scenarios, we have studied the solutions of Bianchi identities in the type IIA and type IIB theories as the same is crucial for finding a genuinely effective scalar potential. In this context we present a peculiar class of solutions, what we call as the `special solutions' of Bianchi identities, in each of the two type II theories. The main idea behind the existence of such solutions is the fact that several Bianchi identities can be understood as a set of orthogonal symplectic (flux) vectors and hence half of the flux components can be rotated away by a symplectic transformation. The possible non-zero fluxes for the `special solutions' are summarized in table \ref{tab_special-flux-sol}. Moreover, after exploring the $T$-dual versions of these `special solutions' from type IIA to type IIB and vice-versa, we make some very interesting observations as collected in the following points:
\begin{itemize}
\item{The non-geometric Type IIA setup with the `special solutions' of Bianchi identities is equivalent to type IIB setup without any  non-geometric fluxes. Moreover, for such a type IIB geometric setup with $O3/O7$, there is a de-Sitter no-go theorem \cite{Shiu:2011zt, Garg:2018reu}, which we have also re-derived from our approach. This helps us in concluding that the $T$-dual type IIA setting which although includes some non-geometric fluxes, cannot result in stable de-Sitter vacua, and this is something against the naive expectations.} 
\item{The non-geometric Type IIB setup with `special solutions' of Bianchi identities is equivalent to type IIA setup without any non-geometric fluxes turned-on. Such a type IIA setup has been studied in a variety of models in the past, especially regarding the search of de-Sitter vacua and their no-go conditions \cite{Hertzberg:2007wc, Flauger:2008ad, Haque:2008jz, Banlaki:2018ayh}.}
\end{itemize}
\noindent
In this context of type IIA orientifold compactifications with geometric flux, first we have re-derived several de-Sitter no-go scenarios of \cite{Hertzberg:2007wc, Flauger:2008ad} and have subsequently explored their $T$-dual counterparts in type IIB theory. In particular we have $T$-dualized three classes of type IIA no-go scenarios which are summarized in table \ref{tab_no-go-fluxTdual}. These can be elaborated as:
\begin{itemize}
\item{no-go-1: Type IIB non-geometric setup with $O3/O7$ and having RR flux $F_3$ along with only the rigid fluxes $H_0, \, \omega_{a0}$ and $\hat{Q}^\alpha{}_0$ cannot give stable de-Sitter vacua.}
\item{no-go-2: Type IIB non-geometric setup with $O3/O7$ and having RR flux $F_3$ along with only the `special solutions' of the NS-NS Bianchi identities cannot give stable de-Sitter vacua unless  the $F^0$ component of the $F_3$ flux is non-zero.}
\item{no-go-3: This no-go scenario is rather a restoration of the no-go-2 itself, in the sense of $F^0$ being zero or non-zero getting irrelevant. This can be done by choosing certain compactification geometries which have factorisation in the complex structure moduli space. To be specific, the violation of no-go-2 via including the non-zero $F^0$ flux (of $F_3$) can be avoided if the type IIB compactification is made on a CY threefold which admits a $K3/{\mathbb T}^4$-fibred mirror Calabi Yau threefold having some specific triple intersection numbers along with the need of setting a couple of fluxes to zero.}
\end{itemize}

\begin{table}[H]
\begin{center}
\begin{tabular}{|c||c||c|c|} 
\hline
Scenarios & & Fluxes in Type IIA   \quad  & \quad Fluxes in Type IIB   \\
& & with $D6/O6$ & with $D3/O3$ and $D7/O7$ \\
\hline
\hline
& & & \\
no-go-1& $F$-term & ${\rm H}_0$,  \quad ${\rm H}_k$, \quad ${\rm H}^\lambda$, & $H_0$, \quad $\omega_{a0}$, \quad $\hat{Q}^\alpha{}_0$, \\
& fluxes  & & \\
& & $e_0$,  \quad $e_a$, \quad $m^a$, \quad $m_0$. & $F_0$,  \quad $F_i$, \quad $F^i$, \quad $- F^0$. \\
& & & \\
\hline
\hline
& & & \\
no-go-2& $F$-term & ${\rm H}_0$,  \quad ${\rm H}_k$, \quad ${\rm H}^\lambda$, & $H_0$, \quad $\omega_{a0}$, \quad $\hat{Q}^\alpha{}_0$, \\
and & fluxes  & & \\
no-go-3& & $w_{a0}$, \quad $w_{ak}$, \quad $w_a{}^\lambda$, & $H_i$, \quad $\omega_{ai}$, \quad $\hat{Q}^\alpha{}_{i}$, \\ 
& & & \\
& & $e_0$,  \quad $e_a$, \quad $m^a$, \quad $m_0$. & $F_0$,  \quad $F_i$, \quad $F^i$, \quad $- F^0$. \\
& & & \\
& $D$-term & $\hat{w}_\alpha{}^0$, \quad $\hat{w}_\alpha{}^k$, \quad $\hat{w}_{\alpha \lambda}$, & $-\,R_K$, \quad $-\,Q^a{}_K$, \quad $\hat{\omega}_{\alpha K}$,\\
& fluxes & & \\
\hline
& & & \\
no-scale-& $F$-term & ${\rm H}_0$,  \, \, $w_{a0}$, \,\, ${\rm Q}^a{}_0$, \, \, ${\rm R}_0$, & $H_0$, \, \, $H_i$, \, \, $H^i$, \, \, $- H^0$, \\
structure& fluxes  & & \\
in IIB & & $e_0$,  \quad $e_a$, \quad $m^a$, \quad $m_0$. & $F_0$,  \quad $F_i$, \quad $F^i$, \quad $- F^0$. \\
\hline
\end{tabular}
\end{center}
\caption{T-dual fluxes relevant for the three no-go scenarios.}
\label{tab_no-go-fluxTdual}
\end{table}
\noindent
Note that in table \ref{tab_no-go-fluxTdual} we have also collected the $T$-dual fluxes corresponding to the type IIB no-scale model which has only the $F_3$ and $H_3$ fluxes. This subsequently shows that in the dual type IIA side, one has all the RR fluxes, and NS-NS fluxes of the `rigid' type only, 
for which we have already shown that a de-Sitter no-go condition exists.

To conclude, we have shown in this analysis how one can engineer a pair of $T$-dual setups in type IIA and type IIB theories in which it may be easier to derive some de-Sitter no-go conditions which can be translated into the mirror side. By considering multiple examples, we have presented a kind of recipe for evading or further restoring the no-go window depending on the various ingredients, including the compactification geometries, one could use. Thus one of the main advantages of this work can also be taken as where to not look for the de-Sitters search, and hence refining the vast non-geometric flux landscape for hunting the de-Sitter vacua. 
Moreover, our analysis can also be extended to utilize/investigate the non-geometric type II models for/against the recently proposed Trans-Planckian Censorship Conjecture (TCC) \cite{Bedroya:2019snp} and also its possible connection with the swampland distance conjecture. We hope to report on (some of) these issues in near future \cite{shukla:2019abc1}.

\section*{Acknowledgments}
I am grateful to Fernando Quevedo for his kind support and encouragements. I would like to thank David Andriot, Erik Plauschinn and Thomas Van Riet for useful discussions and communications. 

\newpage
\appendix
\section{A dictionary for the type II non-geometric flux compactifications}
\label{sec_dictionary}
\begin{table}[H]
\begin{center}
\begin{tabular}{|c||c|c|} 
\hline
& & \\
& Type IIA with $D6/O6$  \quad  & \quad Type IIB with $D3/O3$ and $D7/O7$ \\
& & \\
\hline
\hline
& & \\
$F$-term & ${\rm H}_0$,  \quad ${\rm H}_k$, \quad ${\rm H}^\lambda$, & $H_0$, \quad $\omega_{a0}$, \quad $\hat{Q}^\alpha{}_0$, \\
fluxes  & & \\
& $w_{a0}$, \quad $w_{ak}$, \quad $w_a{}^\lambda$, & $H_i$, \quad $\omega_{ai}$, \quad $\hat{Q}^\alpha{}_{i}$, \\ 
& & \\
 & ${\rm Q}^a{}_0$, \quad ${\rm Q}^a{}_k$, \quad ${\rm Q}^{a \lambda}$, & $H^i$, \quad $\omega_a{}^i$, \quad $\hat{Q}^{\alpha i}$, \\
& & \\
& ${\rm R}_0$,  \quad ${\rm R}_k$, \quad ${\rm R}^\lambda$, & $- H^0$, \quad $- \omega_{a}{}^0$, \quad $- \hat{Q}^{\alpha 0}$, \\
& & \\
& $e_0$,  \quad $e_a$, \quad $m^a$, \quad $m_0$. & $F_0$,  \quad $F_i$, \quad $F^i$, \quad $- F^0$. \\
& & \\
\hline
& & \\
$D$-term & $\hat{w}_\alpha{}^0$, \quad $\hat{w}_\alpha{}^k$, \quad $\hat{w}_{\alpha \lambda}$, & $-\,R_K$, \quad $-\,Q^a{}_K$, \quad $\hat{\omega}_{\alpha K}$,\\
fluxes & & \\
& $\hat{\rm Q}^{\alpha 0}$, \quad  $\hat{\rm Q}^{\alpha k}$, \quad $\hat{\rm Q}^{\alpha}{}_\lambda$. & $-\,R^K$, \quad $-\,Q^{a K}$, \quad $\hat{\omega}_{\alpha}{}^K$.\\
& & \\
\hline
& & \\
Complex & \, \, ${\rm N}^0$, \, \,  ${\rm N}^k$, \, \, ${\rm U}_\lambda$, \, \, ${\rm T}^a$. & $S$, \, \, $G^a$, \, \, $T_\alpha$, \, \, $U^i$.\\
Moduli& & \\
& ${\rm T}^a = \, {\rm b}^a - i\, \, \rmt^a$, & $U^i = v^i  - i \, u^i$,\\
& ${\rm N}^0 = \, \xi^0 + \, i \, ({\rmz}^0)^{-1}$, & $S = c_0 + i\, s\,$, \\
& ${\rm N}^{k} =\, \xi^{k} + \, i \, ({\rmz}^0)^{-1} \, {\rmz}^k$, & $G^a =\left(c^a + c_0 \, b^a \right) + \, i \, s \, \, b^a$,\\
& ${\rm U}_\lambda = -\frac{i}{2\,{\rm z}^0}(k_{\lambda\rho\kappa} {\rmz}^\rho {\rmz}^\kappa - \hat{k}_{\lambda k m} {\rmz}^k {\rmz}^m)$ & $T_\alpha = -\frac{i \, s}{2} \, (\ell_{\alpha\beta\gamma} \, \, t^\beta \, t^\gamma\, - \hat{\ell}_{\alpha a b} \, b^a \, b^b) $ \\
& $+\, \xi_\lambda$. & $+ (c_\alpha +  \hat{\ell}_{\alpha a b} \, c^a b^b + \frac{1}{2} \, c_0 \, \hat{\ell}_{\alpha a b} \, b^a \, b^b)$.\\
& & \\
\hline
& & \\
Axions &  \, \, ${\rm z}^k$, \, \, ${\rm b}^a$, \, \, $\xi^0$, \, \, $\xi^k$, & $b^a$, \, \, $v^i$, \, \, $c_0$,  \, \, $c^a + c_0 \, b^a$, \\
 & $\xi_\lambda$. & $c_\alpha + \hat{\ell}_{\alpha ab}c^a b^b + \frac{1}{2}\, c_0 \, \hat{\ell}_{\alpha a b}b^a b^b$. \\
& & \\
Saxions & $({\rm z}^0)^{-1}$, \, \, ${\rm z}^\lambda$, \, \, $\rmt^a$, \quad ${\cal V}$, \, \, ${\cal U}$, & $s \equiv e^{-\phi}$, \, \, $t^\alpha$, \,\, $u^i$ \quad ${\cal U}$, \, \, ${\cal V}$,\\
 & & \\
\hline
Inter- & & \\
sections & ${k}_{\lambda\rho\mu}\,, \qquad \hat{k}_{\lambda m n} \,, \qquad {\kappa}_{abc} \,, \qquad \hat{\kappa}_{a\alpha\beta}$. & ${\ell}_{\alpha\beta\gamma}, \qquad \hat{\ell}_{\alpha a b}, \qquad {l}_{ijk}, \qquad \hat{l}_{iJK}$. \\
& & \\
\hline
\end{tabular}
\end{center}
\caption{One-to-one T-duality transformations among the various fluxes, moduli and the axions.}
\label{tab_summaryTdual}
\end{table}

\noindent
\begin{table}[H]
\begin{center}
\begin{tabular}{|c||c|} 
\hline
& \\
& Type IIA flux orbits \\
& \\
\hline
\hline
${\rm f}_0 $  & ${\mathbb G}_0 - \, \xi^{\hat{k}} \, {\cal H}_{\hat k} - {\xi}_\lambda \, {\cal H}^\lambda$ \\
${\rm f}_a$  & ${\mathbb G}_a - \, \xi^{\hat{k}} \, {\mho}_{a \hat k} - {\xi}_\lambda \, {\mho}_a{}^\lambda$ \\
${\rm f}^a$  & ${\mathbb G}^a - \, \xi^{\hat{k}} \, {\cal Q}^a{}_{\hat k} - {\xi}_\lambda \, {\cal Q}^{a\lambda}$ \\
${\rm f}^0$ & ${\mathbb G}^0 - \, \xi^{\hat{k}} \, {\cal R}_{\hat k} - {\xi}_\lambda \, {\cal R}^\lambda $ \\
& \\
${\rm h}_0$ & ${\cal H}_0 + {\cal H}_k \, {\rmz^k} \, + \, \frac{1}{2} \, \hat{k}_{\lambda mn} \rmz^m \rmz^n \, {\cal H}^\lambda $ \\
${\rm h}_a$  & ${\cal \mho}_{a0} + {\cal \mho}_{ak} \, {\rmz^k} \, +  \, \frac{1}{2} \, \hat{k}_{\lambda mn} \rmz^m \rmz^n \, {\cal \mho}_a{}^\lambda$ \\
${\rm h}^a$ & ${\cal Q}^a{}_0 + {\cal Q}^a{}_k \, {\rmz^k} \, +  \, \frac{1}{2} \, \hat{k}_{\lambda mn} \rmz^m \rmz^n \, {\cal Q}^{\alpha\lambda}$ \\
${\rm h}^0$ & ${\cal R}_0 + {\cal R}_k \, {\rmz^k} \, + \, \frac{1}{2} \, \hat{k}_{\lambda mn} \rmz^m \rmz^n \, {\cal R}^\lambda $ \\
& \\
${\rm h}_{k0}$ & ${\cal H}_k +  \, \hat{k}_{\lambda k n}\, {\rmz^n} \, {\cal H}^\lambda$ \\
${\rm h}_{ak}$  & ${\cal \mho}_{ak} + \, \hat{k}_{\lambda k n}\, {\rmz^n} \, {\cal \mho}_a{}^\lambda $ \\
${\rm h}^a{}_{k}$ & ${\cal Q}^a{}_k + \, \hat{k}_{\lambda k n}\, {\rmz^n} \, {\cal Q}^{a\lambda}$ \\
${\rm h}_k{}^0$  & ${\cal R}_k +  \, \hat{k}_{\lambda k n}\, {\rmz^n} \, {\cal R}^\lambda $ \\
& \\
${\rm h}^\lambda{}_0$  & ${\cal H}^\lambda$ \\
${\rm h}_a{}^\lambda$  & ${\cal \mho}_a{}^\lambda$ \\
${\rm h}^{a\lambda}$  & ${\cal Q}^{a\lambda}$ \\
${\rm h}^{\lambda 0}$ & ${\cal R}^\lambda$ \\
\hline
\hline
& \\
$F$-term & ${\mathbb G}_0 = \ov{e}_0 + \, {\rm b}^a\, \ov{e}_a + \frac{1}{2} \, \kappa_{abc} \, {\rm b}^a\, {\rm b}^b \,m^c + \frac{1}{6}\, \kappa_{abc}\,  {\rm b}^a \, {\rm b}^b\, {\rm b}^c \, m_0$, \\
fluxes &  ${\mathbb G}_a = \ov{e}_a + \, \kappa_{abc} \,  {\rm b}^b \,m^c + \frac{1}{2}\, \kappa_{abc}\,  {\rm b}^b\, {\rm b}^c \, m_0$, \\
& ${\mathbb G}^a = m^a + m_0\,  {\rm b}^a$, \\
& ${\mathbb G}^0 = m_0$, \\
& \\
& ${\cal  H}_{\hat k} \, \, = \ov{\rm H}_{\hat k} + \ov{w}_{a {\hat k}}\, {\rm b}^a + \frac{1}{2} \kappa_{abc} \, {\rm b}^b \, {\rm b}^c \, {\rm Q}^a{}_{\hat k} + \frac{1}{6} \kappa_{abc} \, {\rm b}^a \, {\rm b}^b \, {\rm b}^c \, {\rm R}_{\hat k}$, \\
& ${\cal  H}^{\lambda} \, \, = \ov{\rm H}^\lambda + \ov{w}_{a}{}^\lambda\, {\rm b}^a + \frac{1}{2} \kappa_{abc} \, {\rm b}^b \, b^c \, {\rm Q}^{a\lambda} + \frac{1}{6} \kappa_{abc} \, {\rm b}^a \, {\rm b}^b \, {\rm b}^c \, {\rm R}^\lambda$, \\
& ${\cal \mho}_{a {\hat k}} = \ov{w}_{a {\hat k}} + \kappa_{abc} \, {\rm b}^b \, {\rm Q}^c{}_{\hat k} + \frac{1}{2} \kappa_{abc} \, {\rm b}^b\, {\rm b}^c \, {\rm R}_{\hat k}$, \\
& ${\cal \mho}_a{}^\lambda = \ov{w}_a{}^\lambda + \kappa_{abc} {\rm b}^b \, {\rm Q}^{c \lambda} + \frac{1}{2} \kappa_{abc} \, {\rm b}^b\, {\rm b}^c \, {\rm R}^\lambda$, \\
& \\
& ${\cal Q}^a{}_{\hat k} = {\rm Q}^a{}_{\hat k} + \, {\rm b}^a\, {\rm R}_{\hat k}$, \quad ${\cal Q}^{a\lambda} = {\rm Q}^{a \lambda}+ \, {\rm b}^a\, {\rm R}^\lambda$, \\
& ${\cal R}_{\hat k} \, \,\,= \, {\rm R}_{\hat k}$, \quad ${\cal R}^\lambda \, \,\,= \, {\rm R}^\lambda$. \\
\hline
& \\
$D$-term & $\hat{\rm h}_{\alpha\lambda} \equiv \hat\mho_{\alpha\lambda} = \hat{w}_{\alpha \lambda} + \hat{k}_{\lambda km} {\rmz}^m \, \hat{w}_\alpha{}^{k} - \frac{1}{2} \hat{k}_{\lambda km} {\rmz}^k {\rmz}^m \, \hat{w}_\alpha{}^{0}$\\
fluxes & $\hat{\rm h}_\alpha{}^{k} \equiv \hat\mho_\alpha{}^{k} = \hat{w}_\alpha{}^{k} - \, {\rm z}^k\, \hat{w}_\alpha{}^{0}$, \quad $\hat{\rm h}_\alpha{}^0 \equiv \hat\mho_\alpha{}^{0} = \, \hat{w}_\alpha{}^{0}$, \\
& \\
& $\hat{\rm h}^{\alpha}{}_{\lambda} \equiv \hat{{\cal Q}}^{\alpha}{}_{\lambda} = \hat{{\rm Q}}^{\alpha}{}_{\lambda} + \hat{k}_{\lambda km} \, {\rmz}^m \, \hat{\rm Q}^{\alpha k}  - \frac{1}{2} \hat{k}_{\lambda km} {\rmz}^\lambda \,{\rmz}^k \,{\rmz}^m \, \hat{Q}^{\alpha 0}$, \\
& $\hat{\rm h}^{\alpha k} \equiv \hat{\cal Q}^{\alpha k} = \hat{\rm Q}^{\alpha k} - {\rm z}^k \, \hat{\rm Q}^{\alpha 0}$, \quad $\hat{\rm h}^{\alpha 0} \equiv \hat{\cal Q}^{\alpha 0} = \hat{\rm Q}^{\alpha 0}$. \\
\hline
\end{tabular}
\end{center}
\caption{Axionic flux orbits for Type IIA side.}
\label{tab_IIA-Fluxorbits}
\end{table}

\begin{table}[H]
\begin{center}
\begin{tabular}{||c|c||c|} 
\hline
& & \\
 & Type IIB flux orbits  & dual Type IIA  \\
& & flux orbits \\
\hline
\hline
& & \\
$f_0 $ & ${\mathbb F}_0 + v^i\, {\mathbb F}_i + \frac{1}{2}\, l_{ijk}\, v^j\, v^k \, {\mathbb F}^i\, - \frac{1}{6}\, l_{ijk}\, v^i \, v^j\, v^k  \, {\mathbb F}^0$ & ${\rm f}_0 $\\
$f_i$ & ${\mathbb F}_i +\, l_{ijk}\, v^j \, {\mathbb F}^k - \frac{1}{2}\, l_{ijk}\, v^j\, v^k \, {\mathbb F}^0$ & ${\rm f}_a$ \\
$f^i$ & ${\mathbb F}^i - v^i \,{ \mathbb F}^0$  & ${\rm f}^a$ \\
$f^0$ & $- \,{\mathbb F}^0$ & ${\rm f}^0$ \\
& & \\
$h_0$ & ${\mathbb H}_0 + v^i\, {\mathbb H}_i + \frac{1}{2}\, l_{ijk}\, v^j\, v^k \, {\mathbb H}^i\, - \frac{1}{6}\, l_{ijk}\, v^i \, v^j\, v^k  \, {\mathbb H}^0$ & ${\rm h}_0$ \\
$h_i$ & ${\mathbb H}_i +\, l_{ijk}\, v^j \, {\mathbb H}^k - \frac{1}{2}\, l_{ijk}\, v^j\, v^k \, {\mathbb H}^0$ & ${\rm h}_a$ \\
$h^i$ & ${\mathbb H}^i - v^i \,{ \mathbb H}^0$ & ${\rm h}^a$ \\
$h^0$ & $-\, {\mathbb H}^0$ & ${\rm h}^0$ \\
& & \\
$h_{a0}$ & ${\mathbb \mho}_{a0} + v^i\, {\mathbb \mho}_{ai} + \frac{1}{2}\, l_{ijk}\, v^j\, v^k \, {\mathbb \mho}_a{}^i\, - \frac{1}{6}\, l_{ijk}\, v^i \, v^j\, v^k  \, {\mathbb \mho}_a{}^0$ & ${\rm h}_{k0}$ \\
$h_{ai}$ & ${\mathbb \mho}_{ai} +\, l_{ijk}\, v^j \, {\mathbb \mho}_a{}^k - \frac{1}{2}\, l_{ijk}\, v^j\, v^k \, {\mathbb \mho}_a{}^0 $ & ${\rm h}_{ak}$ \\
$h_{a}{}^i$ & ${\mathbb \mho}_a{}^i - v^i \,{ \mathbb \mho}_a{}^0$ & ${\rm h}^{a}{}_k$ \\
$h_a{}^0$ & $-\, {\mathbb \mho}_a{}^0 $ & ${\rm h}_k{}^0$ \\
& & \\
$h^\alpha{}_0$ & $\hat{\mathbb Q}_0{}^\alpha + v^i\, \hat{\mathbb Q}_i{}^\alpha + \frac{1}{2}\, l_{ijk}\, v^j\, v^k \, \hat{\mathbb Q}^{\alpha i}\, - \frac{1}{6}\, l_{ijk}\, v^i \, v^j\, v^k  \, \hat{\mathbb Q}^{\alpha 0}$ & ${\rm h}^\lambda{}_0$ \\
$h^\alpha{}_i$ & $\hat{\mathbb Q}_i{}^\alpha +\, l_{ijk}\, v^j \, \hat{\mathbb Q}^{\alpha k} - \frac{1}{2}\, l_{ijk}\, v^j\, v^k \, \hat{\mathbb Q}^{\alpha 0}$  & ${\rm h}_a{}^\lambda$ \\
$h^{\alpha i}$ & $\hat{\mathbb Q}^{\alpha i} - v^i \,\hat{ \mathbb Q}^{\alpha 0}$ & ${\rm h}^{a \lambda}$ \\
$h^{\alpha 0}$ & $-\, \hat{\mathbb Q}^{\alpha 0}$ & ${\rm h}^{\lambda 0}$ \\
\hline
\hline
& & \\
$F$-term & ${\mathbb F}_\Lambda = \ov{F}_\Lambda - \ov{\omega}_{a\Lambda} \, {c}^a - \ov{\hat{Q}^\alpha}{}_\Lambda \,(c_\alpha + \hat{\ell}_{\alpha a b} c^a b^b)   - \, c_0 \, \, {\mathbb H}_\Lambda$ & \\
fluxes & ${\mathbb F}^\Lambda = F^\Lambda - \omega_a{}^\Lambda \, {c}^a - \hat{Q}^{\alpha \Lambda} \, (c_\alpha + \hat{\ell}_{\alpha a b} c^a b^b)\, - \, c_0 \, \, {\mathbb H}^\Lambda$ & \\
& & \\
& ${\mathbb H}_\Lambda = \ov{H}_\Lambda + \ov{\omega}_{a\Lambda} \, {b}^a + \frac{1}{2}\, \hat{\ell}_{\alpha a b}\, b^a b^b \, \ov{\hat{Q}^\alpha}{}_\Lambda$ & \\
& ${\mathbb H}^\Lambda = H^\Lambda + \omega_{a}{}^{\Lambda} \, {b}^a + \frac{1}{2}\, \hat{\ell}_{\alpha a b}\, b^a b^b \, \hat{Q}^{\alpha \Lambda}$ & \\
& ${\mathbb\mho}_{a\Lambda} = \ov{\omega}_{a\Lambda} + \ov{\hat{Q}^\alpha}{}_\Lambda \, \hat{\ell}_{\alpha a b}\, b^b$ & \\
& ${\mathbb\mho}_{a}{}^{\Lambda} = {\omega}_{a}{}^{\Lambda} + \hat{Q}^{\alpha \Lambda} \, \hat{\ell}_{\alpha a b}\, b^b$ & \\
& $\hat{\mathbb Q}^\alpha{}_\Lambda = \ov{\hat{Q}^\alpha}{}_\Lambda, \quad \hat{\mathbb Q}^{\alpha \Lambda} = \hat{Q}^{\alpha \Lambda}$ & \\
& & \\
\hline
& & \\
$D$-term & $\hat{h}_{\alpha K} \equiv \hat{\mho}_{\alpha K} = \hat{\omega}_{\alpha K}\, - Q^{a}{}_{K} \, \hat{\ell}_{\alpha a b} \, b^b +  \frac{1}{2}\hat{\ell}_{\alpha a b} \, b^a \,b^b\, R_K$ & $\hat{\rm h}_{\alpha\lambda}$ \\
fluxes & $\hat{h}_{\alpha}{}^{K} \equiv \hat{\mho}_{\alpha}{}^{K} =\hat{\omega}_{\alpha}{}^{K}\, - Q^{a K} \,\hat{\ell}_{\alpha a b} \, b^b + \frac{1}{2}\hat{\ell}_{\alpha a b} \, b^a \,b^b \, R^K$ & $\hat{\rm h}^{\alpha}{}_{\lambda}$ \\
& & \\
& ${h}^{a}{}_{K}  \equiv {\mathbb Q}^{a}{}_{K} = -{Q}^{a}{}_{K} + R_K b^a, \quad {h}^{a{}K} \equiv {\mathbb Q}^{a{}K} = - {Q}^{a{}K} + R^K b^a$ & $\hat{\rm h}_\alpha{}^k$, \quad $\hat{\rm h}^{\alpha k}$ \\
& & \\
& $\hat{h}_K{}^0 \equiv - {\mathbb R}_K = -\, R_K, \quad \hat{h}^{K0} \equiv - {\mathbb R}^K = -\, R^K$ & $\hat{\rm h}_{\alpha}{}^0$, \quad $\hat{\rm h}^{\alpha 0}$ \\
& & \\
\hline
\end{tabular}
\end{center}
\caption{Axionic Type IIB flux orbits with their dual type IIA counterpart.}
\label{tab_IIB-Fluxorbits}
\end{table}

\noindent
\begin{table}[H]
\begin{center}
\begin{tabular}{|c||c|} 
\hline
& \\
IIA & $V_{\rm IIA}^{\rm tot} = \frac{e^{4D}}{4\, {\cal V}}\biggl[{\rm f}_0^2 + {\cal V}\, {\rm f}^a \, \tilde{\cal G}_{ab} \, {\rm f}^b + {\cal V}\, {\rm f}_a \, \tilde{\cal G}^{ab} \,{\rm f}_b + {\cal V}^2\, ({\rm f}^0)^2\biggr]\, + \frac{e^{2D}}{4\,{\cal U}\,{\cal V}}\biggl[{\rm h}_0^2 + {\cal V}\, {\rm h}^a \, \tilde{\cal G}_{ab} \, {\rm h}^b $ \\
& $ + {\cal V}\, {\rm h}_a \, \tilde{\cal G}^{ab} \,{\rm h}_b + {\cal V}^2\, ({\rm h}^0)^2 + \, {\cal U}\, \tilde{\cal G}^{ij}\,\Bigl({\rm h}_{i0} \, {\rm h}_{j0} + \frac{\kappa_a\, \kappa_b}{4} \,{\rm h}_i{}^a\, {\rm h}_j{}^b  + \, {\rm h}_{ai} \, {\rm h}_{bj} \, {\rm t}^{a}\, {\rm t}^{b} + {\cal V}^2\, {\rm h}_i{}^0\, {\rm h}_j{}^0 $ \\
& $ - \, \frac{\kappa_a}{2}\, {\rm h}^a{}_i\, {\rm h}_{j0} - \frac{\kappa_a}{2} \,{\rm h}_{i0}\, {\rm h}^a{}_j - {\cal V} \, {\rm t}^a \, {\rm h}_i{}^0 \, {\rm h}_{aj} - {\cal V} \, {\rm t}^a \, {\rm h}_{ai}\, h_j{}^0  \Bigr) + \, {\cal U} \,\tilde{\cal G}_{\lambda \rho} \Bigl({\rm h}^\lambda{}_0 \, {\rm h}^\rho{}_0 + \frac{\kappa_a\, \kappa_b}{4} \, {\rm h}^{\lambda{}a} \, {\rm h}^{\rho{}b} $ \\
& $+ \, {\rm t}^a\, {\rm t}^b\, {\rm h}_a{}^\lambda\, {\rm h}_b{}^\rho + {\cal V}^2 \, {\rm h}^{\lambda0}\, {\rm h}^{\rho 0}  - \, \frac{\kappa_a}{2}\, {\rm h}^\lambda{}_0 \, {\rm h}^{\rho{}a} - \frac{\kappa_a}{2}\, {\rm h}^{\lambda a}\, {\rm h}^\rho{}_0 - {\cal V} \, {\rm t}^a \, {\rm h}^\lambda{}^0\, {\rm h}_a{}^\rho - {\cal V} \, {\rm t}^a \, {\rm h}_a{}^\lambda\, {\rm h}^\rho{}^0  \Bigr) $ \\
& $ + \, \frac{k_\lambda \, k_\rho}{4} \Bigl({\cal V}\, {\rm h}^{a \lambda} \, \tilde{\cal G}_{ab} \, {\rm h}^{b \rho} + {\cal V}\, {\rm h}_a{}^\lambda \, \tilde{\cal G}^{ab} \,{\rm h}_b{}^\rho + {\cal V} \, {\rm t}^a \, {\rm h}^{\lambda{}0}\, {\rm h}_a{}^\rho + {\cal V} \, {\rm t}^a \, {\rm h}_a{}^\lambda\, {\rm h}^{\rho 0} - \, {\rm t}^a\, {\rm t}^b\, {\rm h}_a{}^\lambda\, {\rm h}_b{}^\rho$ \\
& $  + \frac{\kappa_a}{2}\, {\rm h}^\lambda{}_0 \, {\rm h}^{a \rho} + \frac{\kappa_a}{2}\, {\rm h}^{a \lambda} \, {\rm h}^\beta{}_0 - \frac{\kappa_a\, \kappa_b}{4} \, {\rm h}^{a \lambda} \, {\rm h}^{b \rho} \Bigr) - \, 2 \times \frac{k_\lambda}{2} \Bigl({\cal V}\, {\rm h}^a \, \tilde{\cal G}_{ab} \, {\rm h}^{b \lambda} + {\cal V}\, {\rm h}_a \, \tilde{\cal G}^{ab} \,{\rm h}_b{}^\lambda $ \\
& $   + {\cal V} \, {\rm t}^a \, {\rm h}^0\, {\rm h}_a{}^\lambda + {\cal V} \, {\rm t}^a \, {\rm h}_a\, {\rm h}^{\lambda 0} - \, {\rm t}^a\, {\rm t}^b\, {\rm h}_a\, {\rm h}_b{}^\lambda +  \frac{\kappa_a}{2}\, {\rm h}^a\, {\rm h}_0{}^\lambda + \frac{\kappa_a}{2}\, {\rm h}_0 \, {\rm h}^{a\lambda} - \frac{\kappa_a\, \kappa_b}{4} \, {\rm h}^a \, {\rm h}^{b \lambda}\Bigr)$ \\
& $+\left[({\cal U} \, \hat{\rm h}_\alpha{}^{0} + {\rmz}^\lambda \, \hat{\rm h}_{\alpha \lambda}) \,{\cal V}\tilde{\cal G}^{\alpha\beta} \,({\cal U} \, \hat{\rm h}_\beta{}^{0} + {\rmz}^\rho \, \hat{\rm h}_{\beta \rho}) + ({\cal U} \, \hat{\rm h}^{\alpha 0} + {\rmz}^\lambda \, \hat{\rm h}^{\alpha}{}_{\lambda}) \,{\cal V}\tilde{\cal G}_{\alpha\beta} \,({\cal U} \, {\rm h}^{\beta 0} + {\rmz}^\rho \, \hat{\rm h}^{\beta}{}_{\rho}) \right] \biggr] $ \\
& $ + \frac{e^{3D}}{2\, \sqrt{\cal U}} \biggl[\left({\rm f}^0 \, {\rm h}_0 - {\rm f}^a\, {\rm h}_a + {\rm f}_a\, {\rm h}^a - {\rm f}_0\, {\rm h}^0 \right) - \left({\rm f}^0\, {\rm h}^\lambda{}_0 - {\rm f}^a\, {\rm h}^\lambda{}_a + {\rm f}_a\, {\rm h}^{\lambda a} - {\rm f}_0\, {\rm h}^{\lambda 0} \right)\, \frac{k_\lambda}{2} \biggr]$.\\
& \\
& $\tilde{\cal G}_{ab} = \frac{\kappa_a\, \kappa_b - 4\, {\cal V}\, \kappa_{ab}}{4\,{\cal V}}, \quad \tilde{\cal G}^{ab} =  \frac{2\, {\rm t}^a \, {\rm t}^b - 4\, {\cal V}\, \kappa^{ab}}{4\,{\cal V}}, \quad \tilde{\cal G}_{\alpha\beta} = -\, \hat{\kappa}_{\alpha\beta}, \quad \tilde {\cal G}^{\alpha\beta} = -\, \hat{\kappa}^{\alpha\beta},$ \\
& \\
& $\tilde{\cal G}_{\lambda\rho} = \frac{k_\lambda\, k_\rho - 4\, {\cal U}\, k_{\lambda\rho}}{4\,{\cal U}}, \quad \tilde{\cal G}^{\lambda\rho} =  \frac{2\, {\rm z}^\lambda \, {\rm z}^\rho - 4\, {\cal U}\, k^{\lambda\rho}}{4\,{\cal U}}, \quad \tilde {\cal G}_{jk} = -\, \hat{k}_{jk}, \quad \tilde {\cal G}^{jk} = -\, \hat{k}^{jk}$.\\
& \\
\hline
& \\
IIB & $V_{\rm IIB}^{\rm tot} = \frac{e^{4\phi}}{4{\cal V}^2\, {\cal U}}\biggl[f_0^2 + {\cal U}\, f^i \, {\cal G}_{ij} \, f^j + {\cal U}\, f_i \, {\cal G}^{ij} \,f_j + {\cal U}^2\, (f^0)^2\biggr]\, + \frac{e^{2\phi}}{4{\cal V}^2\,{\cal U}} \biggl[h_0^2 + {\cal U}\, h^i \, {\cal G}_{ij} \, h^j $ \\
& $ + \, {\cal U}\, h_i \, {\cal G}^{ij} \,h_j + {\cal U}^2\, (h^0)^2 + \, {\cal V}\, {\cal G}^{ab}\,\Bigl(h_{a0} \, h_{b0} + \frac{l_i\, l_j}{4} \,h_a{}^i\, h_b{}^j  + \, h_{ai} \, h_{bj} \, u^{i}\, u^{j} + {\cal U}^2\, h_a{}^0\, h_b{}^0 $ \\
& $ - \, \frac{l_i}{2}\, h_a{}^i\, h_{b0} - \frac{l_i}{2} \,h_{a0}\, h_b{}^i - {\cal U} \, u^i \, h_a{}^0 \, h_{bi} - {\cal U} \, u^i \, h_b{}^0 \, h_{ai}  \Bigr) + \, {\cal V} \,{\cal G}_{\alpha \beta}\,\Bigl(h^\alpha{}_0 \, h^\beta{}_0 + \frac{l_i\, l_j}{4} \, h^\alpha{}^i \, h^\beta{}^j $ \\
& $  + \, u^i\, u^j\, h^\alpha{}_i\, h^\beta{}_j + {\cal U}^2 \, h^\alpha{}^0\, h^\beta{}^0 - \, \frac{l_i}{2}\, h^\alpha{}_0 \, h^\beta{}^i - \frac{l_i}{2}\, h^\alpha{}^i \, h^\beta{}_0 - {\cal U} \, u^i \, h^\alpha{}^0\, h^\beta{}_i - {\cal U} \, u^i \, h^\alpha{}_i\, h^\beta{}^0 \Bigr) $ \\
& $ + \, \frac{\ell_\alpha \, \ell_\beta}{4} \Bigl({\cal U}\, h^\alpha{}^i \, {\cal G}_{ij} \, h^{\beta j} + {\cal U}\, h^\alpha{}_i \, {\cal G}^{ij} \,h^\beta{}_j + {\cal U} \, u^i \, h^\alpha{}^0\, h_i{}^\beta + {\cal U} \, u^i \, h^\alpha{}_i\, h^\beta{}^0 - \, u^i\, u^j\, h^\alpha{}_i\, h^\beta{}_j $ \\
& $ + \frac{l_i}{2}\, h^\alpha{}_0 \, h^\beta{}^i + \frac{l_i}{2}\, h^\alpha{}^i \, h^\beta{}_0 - \frac{l_i\, l_j}{4} \, h^\alpha{}^i \, h^\beta{}^j \Bigr) - \, 2 \times \frac{\ell_\alpha}{2} \Bigl({\cal U}\, h^i \, {\cal G}_{ij} \, h^{\alpha j} + {\cal U}\, h_i \, {\cal G}^{ij} \,h^\alpha{}_j $ \\
& $ + \, {\cal U} \, u^i \, h^0\, h^\alpha{}_i + {\cal U} \, u^i \, h_i\, h^\alpha{}^0 - \, u^i\, u^j\, h_i\, h^\alpha{}_j +  \frac{l_i}{2}\, h^i\, h^\alpha{}_0 + \frac{l_i}{2}\, h_0 \, h^\alpha{}^i - \frac{l_i\, l_j}{4} \, h^i \, h^\alpha{}^j \Bigr) $ \\
& $ +\left[({\cal V} \hat{h}_J{}^{0} - {t}^\alpha \hat{h}_{\alpha J}) {\cal U}{\cal G}^{JK} ({\cal V} \hat{h}_K{}^{0} - {t}^\beta \hat{h}_{\beta K}) + ({\cal V} \, \hat{h}^{J0} - {t}^\alpha \hat{h}_\alpha{}^J) {\cal U}{\cal G}_{JK} ({\cal V} \hat{h}^{K0} - {t}^\beta  \hat{h}_{\beta}{}^K) \right] \biggr]\, $ \\
& $ + \frac{e^{3\phi}}{2{\cal V}^2} \, \biggl[\left(f^0 \, h_0 - f^i\, h_i + f_i\, h^i - f_0\, h^0 \right)\, - \left(f^0\, h^\alpha{}_0 - f^i\, h^\alpha{}_i + f_i\, h^{\alpha i} - f_0\, h^{\alpha 0} \right)\, \frac{\ell_\alpha}{2} \biggr]$. \\
& \\
& $ {\cal G}_{\alpha\beta} = \frac{\ell_\alpha\, \ell_\beta - 4\, {\cal V}\, \ell_{\alpha\beta}}{4\,{\cal V}}, \quad {\cal G}^{\alpha\beta} = \frac{2\, t^\alpha t^\beta - 4\, {\cal V}\, \ell^{\alpha\beta}}{4\,{\cal V}}, \quad {\cal G}_{ab} = -\, \hat{\ell}_{ab},\quad {\cal G}^{ab} = -\, \hat{\ell}^{ab},$\\
& \\
& ${\cal G}_{ij} = \frac{l_i\, l_j - 4\, {\cal U}\, l_{ij}}{4\,{\cal U}}, \quad {\cal G}^{ij} =  \frac{2\, u^i \, u^j - 4\, {\cal U}\, l^{ij}}{4\,{\cal U}}, \quad {\cal G}^{JK} = -\, \hat{l}^{JK}, \quad {\cal G}^{JK} = -\, \hat{l}^{JK}$.\\
& \\
\hline
\end{tabular}
\end{center}
\caption{A one-to-one exchange of the type IIA and type IIB scalar potentials under $T$-duality.}
\label{tab_scalar-potential}
\end{table}


\newpage
\bibliographystyle{utphys}
\bibliography{reference}

\end{document}